\documentclass[twocolumn]{aastex631}
\usepackage{color}
\usepackage[utf8]{inputenc}
\usepackage[T1]{fontenc}
\usepackage{amsmath}

\shortauthors{Zhu et al.}

\begin{document}
	
\title{A New Photoionization Model of the Narrow Line Region in Active Galactic Nuclei}

\author[0000-0002-1333-147X]{Peixin Zhu}
\affiliation{Research School of Astronomy and Astrophysics, Australian National University, Australia}
\affiliation{ARC Centre of Excellence for All Sky Astrophysics in 3 Dimensions (ASTRO 3D), Australia}
\affiliation{Center for Astrophysics $|$ Harvard \& Smithsonian, 60 Garden Street, Cambridge, MA 02138, USA}

\author[0000-0001-8152-3943]{Lisa J. Kewley}
\affiliation{Research School of Astronomy and Astrophysics, Australian National University, Australia}
\affiliation{ARC Centre of Excellence for All Sky Astrophysics in 3 Dimensions (ASTRO 3D), Australia}
\affiliation{Center for Astrophysics $|$ Harvard \& Smithsonian, 60 Garden Street, Cambridge, MA 02138, USA}

\author[0000-0002-6620-7421]{Ralph S. Sutherland}
\affiliation{Research School of Astronomy and Astrophysics, Australian National University, Australia}
\email{peixin.zhu@cfa.harvard.edu}

\begin{abstract}

The photoionization model of narrow-line regions (NLRs) in active galactic nuclei (AGNs) has been investigated for decades. Many published models are restricted to simple linear scaling abundance relations, dust-free assumption, uniform AGN radiation field, and using one specific photoionization code, which restricts them from providing a satisfactory prediction on a broad range of AGN observations. Through a comprehensive investigation, here we present how the choice of abundance scaling relations, dust inclusion, AGN radiation fields, and different photoionization codes CLOUDY and MAPPINGS affect the predictions on the strength of strong UV, optical, and infrared emission lines. We find the dust-depleted radiation pressure-dominated AGN model built with the latest non-linear abundance sets and photoionization code MAPPINGS V are consistent with AGN observations across a broad range of wavelengths. We also assess new potential HII-AGN separation diagrams in the optical and UV wavelengths. 

\end{abstract}

\keywords{galaxies: active --- galaxies: ISM --- galaxies: Seyfert ---ISM: abundances --- quasars: emission lines}

\section{Introduction}

Spectroscopic observations of active galactic nuclei (AGNs) can provide vital insights into physical processes within or in the vicinity of AGNs because the radiation field from central AGNs excites and ionized the nearby interstellar medium, which then emits emission lines across the electromagnetic spectrum. 

When combined with theoretical AGN models, the emission lines of the narrow line region in AGNs can not only place constraints on the chemical abundance and ionization parameter of the interstellar medium but also enable the study of chemical evolution. For example, \citet{storchi-bergmann_chemical_1998} find two metallicity calibrations from AGN photoionization models, which allow the derivation of the chemical abundance of the NLR gas using two easily observed optical emission line ratios [N~{\sc ii}]$\,\lambda\lambda\,$6548,6584/H$\alpha$ and [O~{\sc iii}]$\,\lambda\lambda\,$4959,5007/H$\beta$ or [O~{\sc ii}]$\,\lambda\,$3727/[O~{\sc iii}]$\,\lambda\lambda\,$4959,5007. Recently, attempts have also been made to simultaneously estimate the gas metallicity and ionization parameter by applying a Bayesian approach based on fitting AGN models to the observed emission line spectra of AGNs \citep{thomas_interrogating_2018, perez-montero_bayesian-like_2019}. 

At UV wavelengths, \citet{nagao_gas_2006} show that the gas metallicity of faint high$-z$ galaxies can be estimated by comparing the observed UV emission line ratios in the NLR of high$-z$ radio galaxies with the prediction from AGN models.

Given the potential power for interpreting data with theoretical AGN models, attempts to model the narrow-line region of AGNs have lasted for decades. Beginning with models of \citet{ferland_are_1983} and \citet{halpern_low_1983}, the consensus has been made that photoionization-dominated AGN models are most successful in reproducing the observed strong line ratios in the narrow-line region \citep[e.g.][]{stasinska_confrontation_1984,binette_ionizing_1988}. This conclusion was later reinforced by the conical shape observed in the NLR in several Seyfert galaxies \citep[e.g.][]{schmitt_anisotropic_1994}. 

To generate the photoionization model for the narrow-line region of AGNs, an ionizing radiation field is required in conjunction with a detailed self-consistent photoionization model such as MAPPINGS \citep[e.g.][]{sutherland_cooling_1993,sutherland_mappings_2018}  or CLOUDY \citep[e.g.][]{ferland_2013_2013,ferland_2017_2017}. In addition, settings on the dust physics, thermodynamics structure, and the gas phase abundance of various elements are also required. However, modelers have debated whether dust should be included in AGN models. Although evidence of the existence of dust has been found in the vicinity of AGN \citep{laor_spectroscopic_1993,radomski_resolved_2003}, \citet{nagao_iron_2003,nagao_gas_2006} and \citet{matsuoka_chemical_2009} claim that dust-free AGN models provide more accurate predictions compared to the dusty model based on the observations of UV emission lines and two optical high-ionization forbidden emission lines of Seyfert 2 galaxies. On the contrary, \citet{dopita_are_2002} and \citet{groves_dusty_2004-1} argue that in terms of their model settings, dusty models are more successful in reproducing the observations in narrow-line regions.

The abundance sets modelers used to adopt in AGN models are also being challenged. Assuming the interstellar medium (ISM) in the NLRs and HII regions share a uniform origin, the abundance sets used in the AGN models are the same as in the HII models, whose traditional treatment is to use the solar abundance set as a reference and linearly scale the abundance of most elements according to the total metallicity to solar metallicity ratio $Z/Z_{\odot}$, with two non-linear scaling exceptions for He and N \citep[e.g.][]{storchi-bergmann_chemical_1998,castro_new_2017}. However, it is found in a recent update on the abundance sets for local star-forming galaxies that the scaling relations for most elements are best fitted by a suite of non-uniform relations, which, when applied to HII region models, significantly improves the predictions for the observed emission-line ratios for star-forming galaxies \citep{2017MNRAS.466.4403N}. 

Conflict in the growth of nitrogen abundance is even more considerable. Generally, two sources contribute to the growth of nitrogen abundance as total metallicity accumulates. While the growth rate from primary nucleosynthesis is generally agreed to be constant, the growth rate from secondary nucleosynthetic sources can vary up to 0.3 dex among different galaxies \citep{gutkin_modelling_2016,dors_new_2017}. As a consequence, the discrepancy on the frequently used theoretical [N~{\sc ii}]$\,\lambda\,$6584/[O~{\sc ii}]$\,\lambda\,$3727 ratio can be as large as 0.5 dex in a high metallicity medium ($12+\log(\rm O/H)\approx9.0$) \citep{groves_dusty_2004-1,castro_new_2017}. 

Disagreement also exists in the selection of density structure in the AGN model. Seyfert galaxies are found to locate closely in some of the optical line ratio diagrams, which in terms of the photoionization model, implies that the ionization parameter U in the narrow-line regions are very similar among AGNs \citep{veron-cetty_emission_2000,dopita_are_2002,castro_new_2017}. Some modelers interpret this phenomenon as an implication for the gas density in the ionized clouds decreasing precisely as a function of the inverse square of the distance from the nucleus center and therefore adopt the density-bounded isochoric (constant density) structure in AGN models \citep[e.g.][]{storchi-bergmann_chemical_1998, nagao_gas_2006, castro_new_2017,perez-montero_bayesian-like_2019}. On the other hand, \citet{dopita_are_2002} demonstrated that by including radiation pressure, radiation-bounded AGN models using isobaric (constant pressure) structure could reproduce this phenomenon and is more realistic than using isochoric structure in AGN models \citep{groves_dusty_2004-1}.

To tackle the contradictions in the settings of dust physics, gas abundance sets, and density structure, detailed comparisons between AGN models with the same settings for all but one factor are required. However, the published AGN models are difficult to compare, because they also differ in the radiation field and the photoionization code they are calculated with. For instance, regarding the photoionization code, some researchers calculate their AGN models using CLOUDY \citep[e.g.][]{storchi-bergmann_chemical_1998,nagao_gas_2006,dors_central_2015,perez-montero_bayesian-like_2019}, while others build their AGN models with MAPPINGS \citep[e.g.][]{dopita_are_2002,groves_dusty_2004-1,thomas_interrogating_2018}. The various versions of CLOUDY and MAPPINGS have developed differently over time, with different atomic data sets and calculation methods. It has not yet been clarified how much impact will be introduced into the AGN model by simply changing the photoionization code. Additionally, most published AGN models are tested with small samples, which may not be able to provide sufficient constraints to resolve the ongoing controversies.

In this paper, we perform a comprehensive investigation to separate and quantify the effects of dust physics, gas abundance sets, and density structure on the theoretical emission lines of AGN models. Our goal is to find reliable and self-consistent sets for the AGN model that can account for the emission-line spectra observed in the narrow line regions. We find that the discrepancy in the theoretical emission lines is generally less than 0.1 dex between MAPPINGS and CLOUDY when using the same inputs. We explore AGN models with different radiation fields to account for the influence brought by the change in the AGN radiation field. With the observational constraints from the spectra of the narrow-line region in optical, UV, and infrared wavelength, we show that the dust-depleted AGN model with radiation pressure-dominated isobaric structure provides the best empirical fit to the distribution of Seyfert 2 galaxies on the line ratio diagnostic diagrams in both optical, UV, and infrared wavelength. 

This paper is structured as follows. The observational sample we adopted for model testing is described in Section~\ref{sec:sample}. The radiation field models of AGNs are introduced in Section~\ref{sec:model}, followed by the photoionization codes used in the AGN model calculation. In Section~\ref{sec:model_test}, we quantify the isolated effects of four factors that primarily impact the prediction of AGN models. In Section~\ref{sec:dia_diag}, we explore our selected AGN model further in the optical, UV, and infrared diagnostic diagrams. Finally, we discuss our results in Section~\ref{sec:disc}. 
 
\section{sample selection}\label{sec:sample}

We collect spectroscopic data for AGNs in optical, UV, mid-IR, and far-IR bands to enable model testing across a broad wavelength range. The Sloan Digital Sky Survey (SDSS) provides the largest spectroscopic data set for nearby galaxies in the optical band. Unlike the rich database in optical wavelength ($\sim2000$ Seyferts), public spectroscopic observations for Seyfert 2 galaxies in UV and IR wavelengths are still limited ($\sim70$ in UV and $\sim50$ in IR). When available, we also include observations of HII regions from the same datasets in the diagrams to indicate their locations.

\subsection{Optical Data}

Our optical sample was collected from Sloan Digital Sky Survey (SDSS) data release 16 (DR16) \citep{kollmeier_sdss-v_2017}, which contains the latest optical emission-line data of a large sample of galaxies. We use SQL to collect the emission-line fluxes of {[O~{\sc ii}]}$\,\lambda\lambda$3726,9, {[Ne~{\sc iii}]}$\,\lambda$3869, H$\beta$, {[O~{\sc iii}]}$\,\lambda\,$5007, H$\alpha$, {[O~{\sc i}]}$\,\lambda\,$6300, [N~{\sc ii}]$\,\lambda$6584, and [S~{\sc ii}]$\,\lambda\,\lambda$6717,31 for galaxies that satisfy the following criteria:

	(a) Signal-to-noise ratio (S/N)$\geq\,$3 in the strong emission-lines H$\beta$, {[O~{\sc iii}]}$\,\lambda\,$5007, H$\alpha$, {[O~{\sc i}]}$\,\lambda\,$6300, [N~{\sc ii}]$\,\lambda$6584, and [S~{\sc ii}]$\,\lambda\,\lambda$6717,31.
	
	(b) Redshifts between $0.05<z<0.1$.
	
	(c)  All emission lines used in this work should have fluxes larger than zero. This is guaranteed by setting `Fit\_Warning'=0. When an emission line falls on the sky line, the `Fit\_Warning' parameter is set to 1, and measurements of the emission line are set to zero.
	
The S/N limit is chosen to ensure that the classification based on strong emission line ratios is accurate and to ensure that there are sufficient galaxies left for model testing \citep[e.g.][]{veilleux_spectral_1987,kewley_host_2006}. The lower redshift is chosen to prevent the galaxy properties from being dominated by the effects of the small fixed size aperture \citep{kewley_aperture_2005}. The upper redshift limit ensures that our sample consists of local galaxies. The third criterion ensures reliable emission line measurements.

Following the steps in \citet{vogt_galaxy_2013}, we performed extinction correction for these optical emission-line fluxes with the Balmer decrement and applied the relative extinction curve with $R^A_V=A_V/E(B-V)=4.5$ and $A_V=1$ of \citet{2005ApJ...619..340F}. We assume an intrinsic H$\alpha$/H$\beta$ ratio of 2.86 for star formation dominated galaxies and H$\alpha$/H$\beta=3.1$ for AGN dominated galaxies. 

With the extinction corrected emission-line fluxes, we further classify these galaxies into star-forming galaxies, AGNs, and composite galaxies on the standard optical diagnostic diagrams \citep{baldwin_classification_1981,veilleux_spectral_1987}, by applying the following classification scheme proposed by \citet{kewley_theoretical_2001,kauffmann_host_2003,kewley_host_2006}. 

	(i) Star-forming galaxies should satisfy equations (1), (2), and (3) on the standard optical diagnostic diagrams.	
	\begin{equation}
	\log(\rm{[O~{\sc III}]}/H\beta) < 0.61 / [\log({[N~{\sc II}]}/H\alpha)-0.05]+1.3 
	\end{equation}
	\begin{equation}
	\log([\rm{O~{\sc III}}]/H\beta)<0.72/[\log({[S~{\sc II}]}/H\alpha)-0.32]+1.3
	\end{equation}
	\begin{equation}
	\log(\rm{O~{\sc III}}]/H\beta)<0.73/[\log([O~{\sc I}]/H\alpha)+0.59]+1.33	
	\end{equation}
	
	(ii)Composite galaxies should satisfy equations (4) and (5) on the {[N~{\sc ii}]}/H$\alpha$ versus {[O~{\sc iii}]}/H$\beta$ diagram.
	\begin{equation}
	\log(\rm{[O~{\sc III}]}/H\beta)>0.61/[\log({[N~{\sc II}]}/H\alpha)-0.05]+1.3
	\end{equation}
	\begin{equation}
	\log(\rm{[O~{\sc III}]}/H\beta)<0.61/[\log({[N~{\sc II}]}/H\alpha)-0.47]+1.19
	\end{equation}
	
	(iii)Seyfert galaxies should satisfy equation (6) on the {[N~{\sc ii}]}/H$\alpha$ versus {[O~{\sc iii}]}/H$\beta$ diagram, equation (7) and (8) on the {[S~{\sc ii}]}/H$\alpha$ versus {[O~{\sc iii}]}/H$\beta$ diagram, and equation (9) and (10) on the {[O~{\sc i}]}/H$\alpha$ versus {[O~{\sc iii}]}/H$\beta$ diagram.
	\begin{equation}
		\log(\rm{[O~{\sc III}]}/H\beta)>0.61/[\log({[N~{\sc II}]}/H\alpha)-0.47]+1.19
	\end{equation}
	\begin{equation}
		\log(\rm{[O~{\sc III}]}/H\beta)>0.72/[\log({[S~{\sc II}]}/H\alpha)-0.32]+1.3 
	\end{equation}
	or $\log(\rm{[S~{\sc II}]}/H\alpha)>0.32$
	\begin{equation}
		\log(\rm{[O~{\sc III}]}/H\beta)>1.89\log([S~{\sc II}]/H\alpha)+0.76
	\end{equation}
	\begin{equation}
	\log(\rm{[O~{\sc III}]}/H\beta)>0.73/[\log([O~{\sc I}]/H\alpha)+0.59]+1.33	
	\end{equation}
	or $\log(\rm[O~{\sc I}]/H\alpha)>-0.59$
	\begin{equation}
	\log(\rm{[O~{\sc III}]}/H\beta)>1.18\log([O~{\sc I}]/H\alpha)+1.30
	\end{equation}

According to the above criteria, our 47302-galaxy sample contains 36243 (77\%) star-forming galaxies, 4946 (10\%) composite galaxies, and 1783 (5\%) Seyfert galaxies. The remaining 3859 galaxies are LINERs and ambiguous galaxies (8\%).

\subsection{UV Data}\label{sec:2.2}

Contrary to optical observations, the available data for AGNs in the UV band is limited and lacks recent measurements. We collect the emission-line fluxes of {[N~{\sc v}]}$\,\lambda$1240, {[He~{\sc ii}]}$\,\lambda$1640, {[C~{\sc iii}]}$\,\lambda$1909, {[C~{\sc iv}]}$\,\lambda$1549 for 77 Type-2 AGN galaxies at redshift $0<z<4.0$ from the Table 1 in \citet{dors_semi-empirical_2019}, who compiled the data from \citet{kraemer_iue_1994,de_breuck_sample_2000,nagao_gas_2006,bornancini_imaging_2007,matsuoka_chemical_2009,matsuoka_mass-metallicity_2018}. These 77 AGN galaxies are selected by applying a S/N$\geq\,$3 limit and consist of nine local Seyfert 2 galaxies ($z<0.03$), seven $1.5<z<3.6$ X-ray selected Type-2 quasars, and 61 high$-z$ radio galaxies (HzRGs) at $1.2<z<4.0$. To account for the dust extinction effect in our comparisons, we adopt the extinction curve described by \citet{1989ApJ...345..245C} and show the $A_V=5.0\,$mag effect in the UV diagnostic diagrams.

\subsection{Mid-IR and Far-IR Data}\label{sec:2.3}

For IR emission lines, we use the IR spectroscopic database presented in Table 3 of \citet{fernandez-ontiveros_far-infrared_2016}. They assembled a sample that contains 57 local Seyfert 2 galaxies ($z<0.45$) and 20 star-forming galaxies ($z<0.0159$) and collected their mid-IR emission lines from the InfraRed Spectrograph (IRS; \citet{houck_infrared_2004}) on board the Spitzer Space Telescope \citep{werner_spitzer_2004} as well as far-IR fine-structure lines from the Spectral and Photometric Imaging Receiver Fourier-transform spectrometer (SPIRE; \citet{griffin_herschel-spire_2010}) on board the Herschel Space Observatory (\citet{pilbratt_herschel_2010}). These 57 Seyfert 2 galaxies are selected from far-IR and mid-IR observations for QSO or Seyferts that have S/N$\geq$3 detections without the presence of broad-lines in optical, IR, and/or polarized light spectra. The 20 star-forming galaxies are selected from optically classified pure HII region galaxies that have mid-IR and far-IR observations with S/N$\geq$3.

We collected the fluxes of {[S~{\sc iv}]}$\,\lambda$10.5$\rm \mu m$, {[Ne~{\sc ii}]}$\,\lambda$12.8$\rm \mu m$, {[Ne~{\sc v}]}$\,\lambda$14.5$\rm \mu m$, {[Ne~{\sc iii}]}$\,\lambda$15.6$\rm \mu m$, {[S~{\sc iii}]}$\,\lambda$18.7$\rm \mu m$, {[Ne~{\sc v}]}$\,\lambda$24.3$\rm \mu m$, {[O~{\sc iv}]}$\,\lambda$25.9$\rm \mu m$, {[S~{\sc iii}]}$\,\lambda$33.5$\rm \mu m$, {[Si~{\sc ii}]}$\,\lambda$34.8$\rm \mu m$ for mid-IR data, and the fluxes of {[O~{\sc iii}]}$\,\lambda$52$\rm \mu m$, {[N~{\sc iii}]}$\,\lambda$57$\rm \mu m$, {[O~{\sc i}]}$\,\lambda$63$\rm \mu m$, {[O~{\sc iii}]}$\,\lambda$88$\rm \mu m$, {[N~{\sc ii}]}$\,\lambda$122$\rm \mu m$, {[O~{\sc i}]}$\,\lambda$145$\rm \mu m$, and {[C~{\sc ii}]}$\,\lambda$158$\rm \mu m$ for far-IR data. Given that the  extinction at $\rm \lambda>3\mu m$ is generally smaller ($A_{\lambda}/A_V<0.1$, \citet{2021ApJ...916...33G}) than the measurement errors of the line fluxes, we did not perform extinction correction for IR data. 

\section{model description}
	\label{sec:model}
	
To build the photoionization model of the narrow-line region in AGNs, the radiation field of AGN is required for calculation in the photoionization code. Here we use the physically based radiation field model OXAF \citep{thomas_physically_2016} to generate the ionizing spectra of AGNs. In this section, we investigate how the change in AGN spectrum shape affects the emission-line spectra of AGN models. For the convenience of external comparison, we also include the CLODUY default AGN radiation field in our experiment, which is widely used in AGN models calculated using CLOUDY.

The most popular two photoionization codes are MAPPINGS \citep[e.g.][]{sutherland_mappings_2018} and CLOUDY \citep[][]{ferland_2017_2017}. As the atomic data and physical process are not the same in these photoionization codes, we also investigate the difference in AGN models introduced by simply changing the photoionization code from the latest version of MAPPINGS (MAPPINGS V) to the 2017 version of CLOUDY (CLOUDY C17), keeping all input parameters and ionizing radiation field constant.

\subsection{Radiation Field}\label{sec:3.1}

\begin{figure*}[htb]
\epsscale{0.8}
\plotone{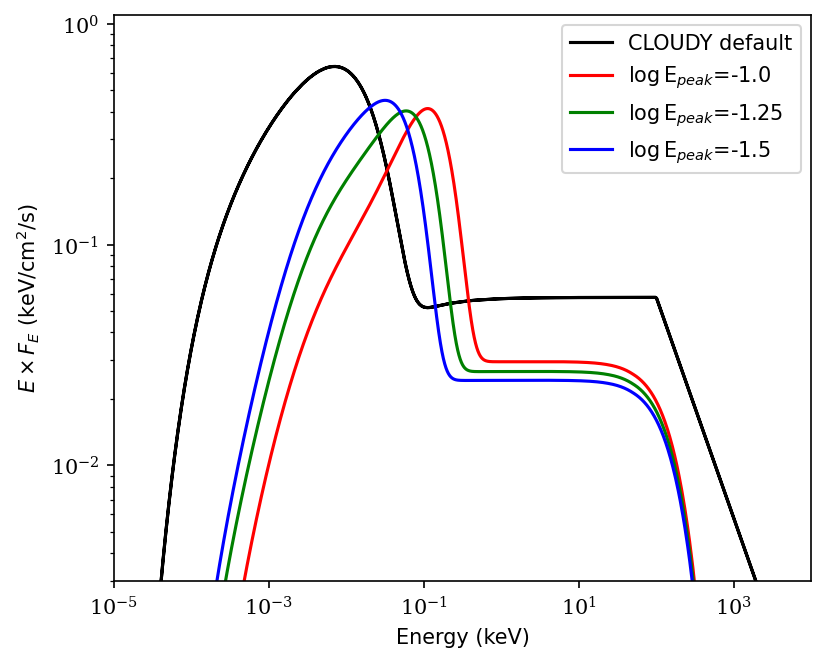}
\caption{Radiation fields of AGN produced by the OXAF model with fixed photo index $\Gamma=2.0$ and the flux ratio of the non-thermal tail $p_{\text{NT}}=0.15$ and peak energy of the accretion disk $\log E_{\text{peak}}/(\text{keV})=-1.5,-1.25,-1.0$, corresponding to $E_{\text{peak}}=32eV, 56eV, 100eV$. The default AGN radiation field provided by CLOUDY is also presented for comparison.
\label{fig:1}}
\end{figure*}

In the past, empirical AGN radiation fields were built with piecewise functions of one to three simple power laws \citep[e.g.][]{viegas-aldrovandi_composite_1989,groves_dusty_2004-1}. Here we adopt a physically based radiation model OXAF introduced and described in detail by \citet{thomas_physically_2016}. Briefly, the OXAF model is a simple version of a more complicated radiation model OPTXAGNF \citep{done_intrinsic_2012,jin_combined_2012-1}, which characterize the essential features of continuum emission radiated from a thin accretion disk with a Comptonizing corona of a rotating black hole with nine parameters. To build the OXAF model using the fewest possible parameters, \citet{thomas_physically_2016} carefully shrank and reparametrized the nine parameters in the OPTXAGNF model and only left three parameters that have the most critical impacts on the NLR heating: the peak energy of the accretion disk emission $E_{\text{peak}}$, the photon index of the inverse Compton scattered power-law tail $\Gamma$, and the flux ratio of the non-thermal tail to the total flux $p_{\text{NT}}$. 

After a systematic investigation, \citet{thomas_physically_2016} found that the theoretical emission line ratios of AGNs are most sensitive to the peak energy of the Big Blue Bump (BBB) disk emission, that is, $E_{\text{peak}}$, while the impact from $\Gamma$ and $p_{\text{NT}}$ are relatively small. Furthermore, $\Gamma$ and $p_{\text{NT}}$ are also found anti-correlated to a certain degree. After investigating a broad parameter space, \citet{thomas_interrogating_2018} suggested that OXAF models with fixed $\Gamma=2.0$ and $p_{\text{NT}}=0.15$ plus a varying $E_{\text{peak}}$ can provide a realistic radiation field for most of the narrow line regions in the vicinity of AGNs. 

Here we compare the shape of different AGN radiation fields. We select four typical AGN radiation fields for our comparison, which consist of three AGN radiation fields from the OXAF model using $\Gamma=2.0$ and $p_{\text{NT}}=0.15$ with $E_{\text{peak}}=32\rm eV, 56eV, 100eV$ and the default AGN radiation field generated by CLOUDY with the command `AGN T=1.5e5k, a(ox)=$-$1.4, a(uv)=$-$0.5 a(x)=$-$1', where T stands for the temperature of the Big Bump continuum, a(ox) controls the X-ray to UV ratio $\alpha_{ox}$, a(uv) represents the low-energy slope of the Big Bump continuum $\alpha_{uv}$, and a(x) reflects the slope of the X-ray component $\alpha_x$ \citep{ferland_2013_2013,ferland_2017_2017}.

As shown in Figure~\ref{fig:1}, the CLOUDY default AGN radiation field has the lowest energy peak in the BBB component, at $\sim1\,$Ryd, while the BBB component moves toward higher energy as the  $E_{\text{peak}}$ in OXAF AGN radiation fields increase. Compared to OXAF radiation fields, the CLOUDY radiation field has a broader BBB component and a smaller slope in the X-ray tail at higher energy. This is because the CLOUDY radiation field combines a single blackbody curve with power law to produce the BBB component and uses a simple piecewise $\nu^{-2}$ cutoff at the high energy tail where the OXAF radiation fields use a power-law cutoff. 

\subsection{Photoionization Codes}	

\begin{figure*}[tb]
\epsscale{1.1}
\plotone{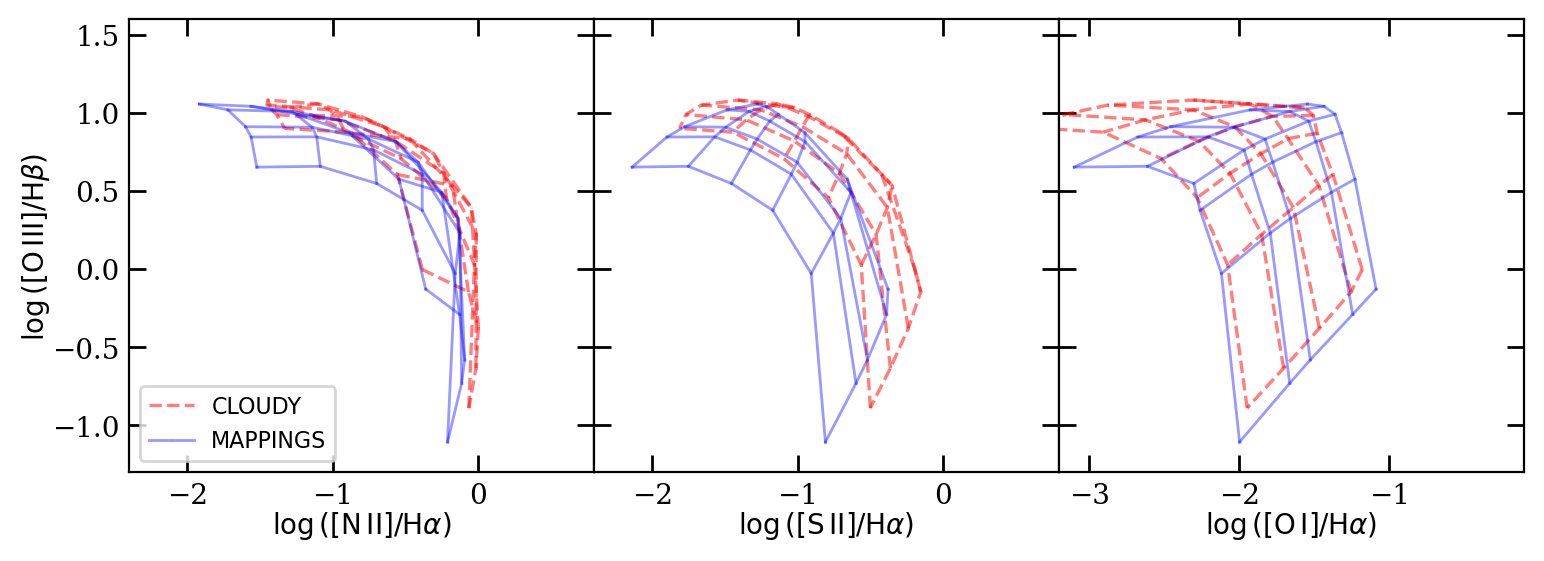}
\caption{Comparison of AGN models calculated by different photoionization codes on the standard optical diagnostic diagrams. Each model is presented with constant metallicity lines (from left to right: $12+\log(\rm O/H)=8.43, 8.70, 8.89, 8.96, 9.09$) and constant ionization parameter lines (from bottom to up: $\log(\rm U)=-3.5, -3.0, -2.5, -2.0, -1.5, -1.0$) at hydrogen density $n_{\rm H}=1000\rm \,cm^{-3}$. The red dashed and solid blue grids represent dust-free, isochoric structure AGNs model with solar-based abundance file and CLOUDY default AGN radiation field calculated by CLOUDY v17 \citep{ferland_2017_2017} and  MAPPINGS V \citep{sutherland_mappings_2018}.
\label{fig:3}}
\end{figure*}

To model the heating and cooling processes that lead to absorption and emission in the interstellar medium, a photoionization code is required. In this work, we adopt the photoionization code MAPPINGS V \citep{sutherland_mappings_2018} to generate the emission-line spectra of AGN as a function of metallicity and ionization parameters. The MAPPINGS V code, described in detail by \citet{sutherland_effects_2017}, is the latest version of the MAPPINGS photoionization code, developed over three decades. Starting from the first iteration \citep{binette_radiative_1985}, the MAPPINGS code was subsequently improved by adding new physical processes including dust heating, infrared emission, and treatment of non-equilibrium electron energies \citep[e.g.][]{sutherland_cooling_1993,dopita_importance_2000,dopita_modeling_2005,dopita_new_2013}. In MAPPINGS V, significant changes include updated atomic data from version 10 of the CHIANTI database \citep{del_zanna_chiantiatomic_2021}, updated elemental depletion files, and higher precision exponential integral functions to calculate collisional ionization rates, which enable MAPPINGS V to track more than $8\times10^4$ cooling and recombination emission lines.

Aside from MAPPINGS, another widely used photoionization code is CLOUDY \citep[e.g.][]{ferland_high_1996}, a microphysics modeling code that was first built in \citet{ferland_are_1983}. We use the 2017 version of CLOUDY (C17) \citep{ferland_2017_2017}, which by default uses the self-built atomic database Stout. Given the differences between MAPPINGS V and C17 in physical processes and atomic data, it is worth investigating how the choice of photoionization code can affect the theoretical emission-line spectra of AGN models. To do this, we provide the same inputs into MAPPINGS and C17, respectively, and compare their predictions in Figure~\ref{fig:3}. Both AGN models are considered dust-free models with isochoric density structure and adopt solar-based abundance and CLOUDY default AGN radiation field.

With the same inputs, MAPPINGS V and C17 give similar predictions in the emission line ratios {[N~{\sc ii}]}/H$\alpha$ and {[O~{\sc i}]}/H$\alpha$, with difference $\Delta$ less than $0.1\,$dex at $\log(U)\lesssim-2.0$. At $\log(U)>-2.0$, the differences rise to $\Delta\approx0.2\,$dex. For {[O~{\sc iii}]}/H$\beta$, the predictions from two photoionization codes are close ($\Delta<0.1\,$dex) for high ionization parameter $\log(U)\gtrsim-3.0$ and relatively low metallicity ($12+\log(\text{O/H})\lesssim8.7$), but differ up to $\Delta\approx0.2\,$dex in regions with lower ionization parameter or higher metallicity. The {[S~{\sc ii}]}/H$\alpha$ ratio suffers the most significant impact from the change of photoionization code, with the difference between the two predictions ranging from $\Delta\approx0.1\,$dex to $\Delta\approx0.3\,$dex. The large discrepancy in the theoretical {[S~{\sc ii}]} flux is expected as the atomic database Stout adopts a $\sim30\%$  different transition energy for SII compared with CHIANTI v10 \citep{kisielius_atomic_2013}.

Overall, the theoretical AGN models calculated by MAPPINGS V and C17 with the same inputs present a similar trend between emission line ratios, gas metallicity, and ionization parameters. Their discrepancies in most theoretical emission line ratios are generally around 0.1 dex, mainly affected by the physical processes used in the photoionization codes. Although theoretical emission lines that are additionally affected by atomic data may have a larger discrepancy (up to $\sim0.3$ dex), we conclude that both photoionization codes, MAPPINGS V and C17, provide similar predictions on the AGN model with the same inputs. The choice of photoionization code does not affect our results in the following section. All AGN models presented in the following sections are calculated by MAPPINGS.

\subsection{The Impact of Ionizing Radiation Fields}

\begin{figure*}[tb]
\epsscale{1.1}
\plotone{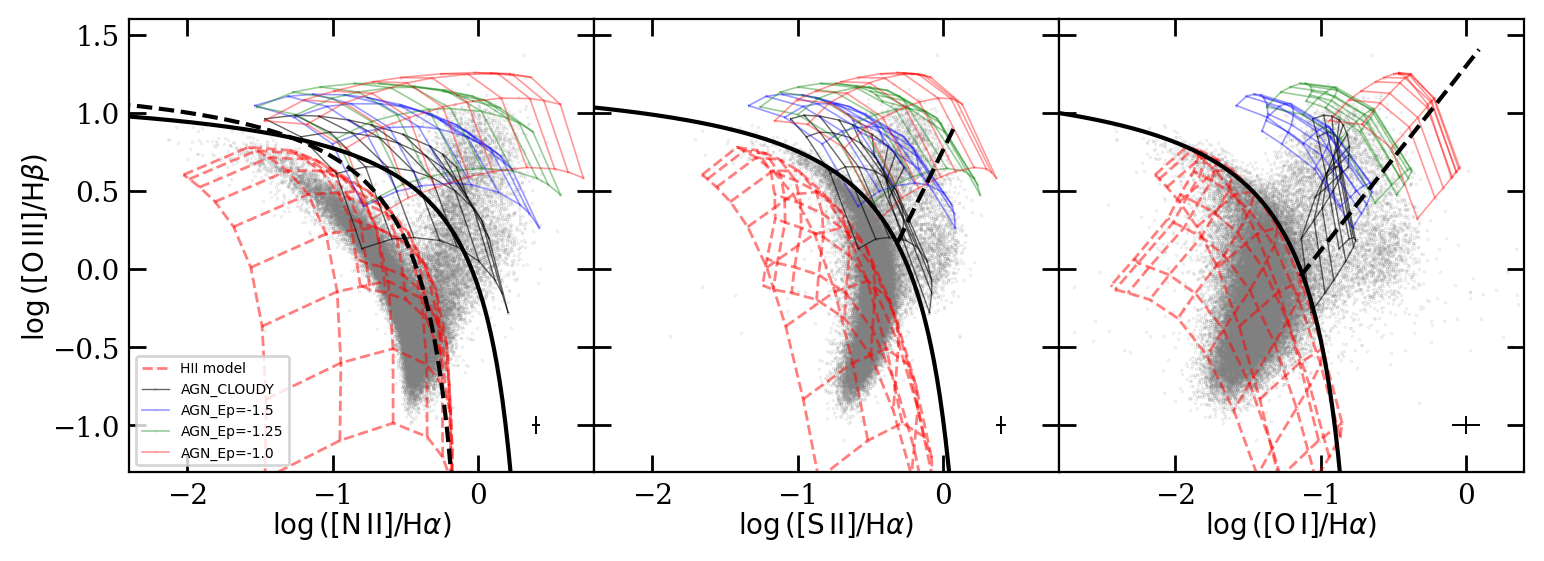}
\caption{Comparison of AGN models built with four different AGN radiation fields on the standard optical diagnostic diagrams. Each model is presented with constant metallicity lines (from left to right: $12+\log(\rm O/H)=8.28, 8.43, 8.54, 8.70, 8.80, 8.96, 9.09, 9.16, 9.26$) and constant ionization parameter lines (from bottom to up: $\log(\rm U)=-3.4, -3.0, -2.6, -2.2$) at gas pressure $\log \rm(P/k)=7.0$. The solid red, green, and blue grids represent AGN models using OXAF ionizing spectra with fixed $\Gamma=2.0$ and $p_{\text{NT}}=0.15$ and $\log E_{\text{peak}}/(\text{keV})=-1.0,-1.25,-1.5$ respectively. The solid black grid shows the AGN model built with the CLOUDY default AGN radiation field. And the Red dashed grid represents the latest version of the HII model (Li et al.2023, in prep) for a complete comparison with the observations. Optical spectra for a large sample of galaxies from SDSS DR16 are presented with grey dots. The mean error bars are shown in the lower right corner. In all three panels, the black solid curves are the \citet{kewley_theoretical_2001} theoretical maximum starburst lines. In the left panel, the black dashed curve is the \citet{kauffmann_host_2003} empirical maximum starburst line. The black dashed lines in the right two panels are the Seyfert-LINER (low-ionization narrow emission-line regions) separation lines \citep{kewley_host_2006}.
\label{fig:2}}
\end{figure*}

To investigate how the shape of the AGN spectrum affects the theoretical AGN models, we compare the AGN models built with different AGN radiation fields on the standard optical diagnostic diagrams. In Figure~\ref{fig:2}, we compare four AGN models constructed with the four AGN radiation fields shown in Figure~\ref{fig:1}. All these models are dusty AGN models built with the non-linear abundance file and isobaric structure with gas pressure $\log \rm(P/k)=7.0$ calculated by MAPPINGS. As shown in Figure~\ref{fig:7} and later discussion, AGN models vary little on the standard optical diagnostic diagrams as gas pressure goes below $\log \rm(P/k)=7.0$. Therefore we use $\log \rm(P/k)=7.0$ as a representative pressure value for AGN models here. We can see from Figure~\ref{fig:2} that the CLOUDY default radiation field produces AGN models that lie lower on the standard optical diagnostic diagrams than the OXAF radiation fields. With the same gas metallicity and ionization parameter, the AGN model that has higher peak energy in the BBB component of the radiation field predicts higher emission line ratios of {[N~{\sc ii}]}/H$\alpha$, {[S~{\sc ii}]}/H$\alpha$, {[O~{\sc i}]}/H$\alpha$, and {[O~{\sc iii}]}/H$\beta$. By comparing the models with the optical spectra of Seyfert galaxies in Figure~\ref{fig:2}, we conclude that $\log E_{\text{peak}}/(\text{keV})=-1.25$ AGN model can adequately account for those Seyfert galaxies that have the highest {[O~{\sc iii}]}/H$\beta$ ratios within a reasonable metallicity and ionization parameter ranges. In contrast, AGN models with lower $E_{\text{peak}}$ in the radiation field fail to predict the behavior of these highly ionized Seyfert galaxies. The model with $\log E_{\text{peak}}/(\text{keV})=-1.0$ overestimates the observed emission line ratios in all nearby Seyfert galaxies. 

As an increase by $0.25\,$dex in the peak energy of the BBB component in AGN radiation field can lead to a $\sim0.3\,$dex enhancement on all four emission line ratios {[N~{\sc ii}]}/H$\alpha$, {[S~{\sc ii}]}/H$\alpha$, {[O~{\sc i}]}/H$\alpha$, and {[O~{\sc iii}]}/H$\beta$, we caveat that the $E_{\text{peak}}$ of the input AGN radiation field into AGN photoionization model should be determined appropriately before applying theoretical metallicity diagnostics for metallicity measurement in AGN regions (Zhu et al. 2023b, in prep).

\section{model testing}\label{sec:model_test}

To model the emission line spectra from an AGN-excited interstellar medium in photoionization models, settings on dust properties, element abundance, density structure, and gas pressure are also required. In this section, we investigate how these four factors affect the behavior of theoretical AGN models on the standard optical diagnostic diagrams and which parameters for each factor can provide the closest prediction to the observed optical spectra of AGNs.

\subsection{Dust Properties}

\begin{figure*}[htb]
\epsscale{0.8}
\plotone{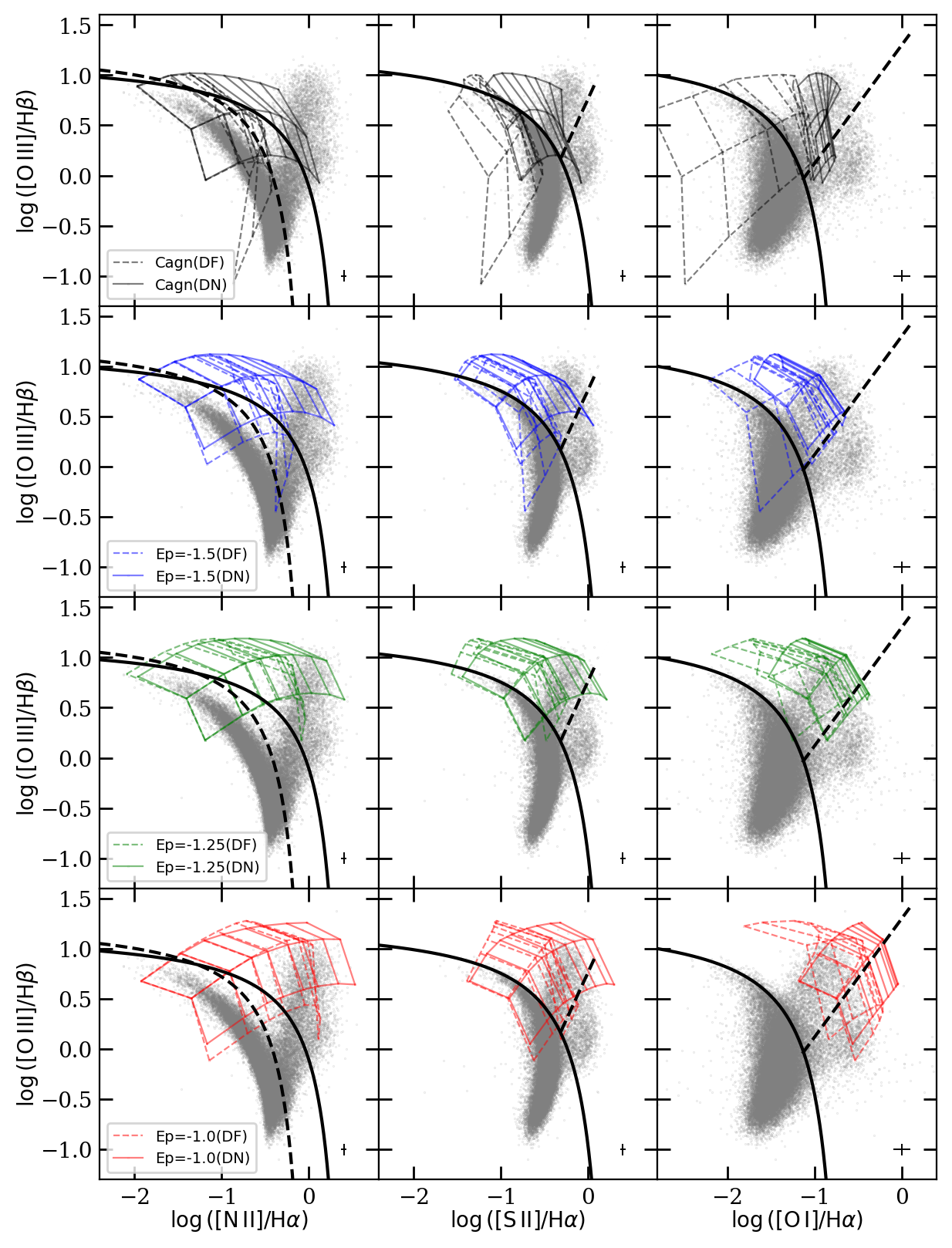}
\caption{Comparison of dust-free (DF) AGN model (dashed lines) and dust-depleted (DN) AGN model (solid lines) on the standard optical diagnostic diagrams. To isolate the effect from the AGN radiation field, four sets of dust-free and dusty AGN models are presented from top to bottom panels using CLOUDY default radiation field (called `Cagn'), OXAF radiation field with $\log E_{\text{peak}}=-1.5, -1.25, -1.0$, respectively. Each model is presented with constant metallicity lines (from left to right: $12+\log(\rm O/H)=8.00, 8.28, 8.43, 8.54, 8.70, 8.80, 8.96, 9.09$) and constant ionization parameter lines (from bottom to up: $\log(\rm U)=-3.4, -3.0, -2.0$) at gas pressure $\log \rm(P/k)=7.0$. Optical spectra for a large sample of galaxies from SDSS DR16 are presented with grey dots. The mean error bars are shown in the lower right corner. In all three panels, the black solid curves are the \citet{kewley_theoretical_2001} theoretical maximum starburst lines. In the left panel, the black dashed curve is the \citet{kauffmann_host_2003} empirical maximum starburst line. The black dashed lines in the right two panels are the Seyfert-LINER (low-ionization narrow emission-line regions) separation lines \citep{kewley_host_2006}.\label{fig:4}}
\end{figure*}

Whether to include dust in the AGN model is a topic that has been debated for over a decade. Supporting evidence for the dust-free model was presented by \citet{nagao_iron_2003} and \citet{nagao_gas_2006}. The former study compared dust-free and dusty AGN models to the observed emission line ratios {[Fe~{\sc vii}]}$\,\lambda\,$6087/{[Ne~{\sc v}]}$\,\lambda\,$3426 in 24 high-ionization Seyfert 2 galaxies ($\log(\text{U})\gtrsim -2.0$), concluding that iron is not depleted by dust in the narrow-line regions of these galaxies, and thus a dust-free model should be preferred. The latter study performed a similar comparison on the UV emission line ratios {[C~{\sc iii}]}$\,\lambda$1909/{[C~{\sc iv}]}$\,\lambda$1549 and {[C~{\sc iv}]}$\,\lambda$1549/{[He~{\sc ii}]}$\,\lambda$1640 observed in the narrow-line regions of 51 high$-z$ radio galaxies, 9 type-2 QSOs, and 9 Seyfert 2 galaxies, finding that AGN model including dust grains fails to predict the majority of observations, unlike the dust-free AGN model. 

The powerful radiation field generated by the AGN has been largely believed to be inimical to the survival of dust in the vicinity of AGN. However, direct evidence of the presence of dust in the narrow-line region of AGNs is found in resolved observations \citep[e.g.][]{laor_spectroscopic_1993,radomski_resolved_2003}. For instance,  \citet{radomski_resolved_2003} studied the resolved mid-infrared image of NGC4151 and found that dust is the most likely explanation for the observed extended mid-IR emission. More recently, a hot dust ring has been discovered in the vicinity of the black hole at a radius that was thought to be evaporated by the central radiation \citep{gamez_rosas_thermal_2022}. Besides, the inclusion of dust is not the only factor that can impact the AGN model. As we show in the following subsections, both abundance scaling relations and density structure can significantly affect predictions of the AGN model.

We carefully compare the dust-free AGN model and the dusty AGN model calculated with MAPPINGS V. We use the same settings on element abundance and density structure to build a dust-free model and a dusty model. We found the combination of non-uniform abundance scaling relations and isobaric density structure provides the closest predictions to observations in the AGN model. In the dusty model, dust formation will deplete the heavy elements from the surrounding gas, especially for those with a higher condensation temperature, and thus change the abundance of heavy elements. The depletion factors are adopted from \citet{thomas_interrogating_2018}, which is derived from \citet{jenkins_unified_2009} where iron is $97.8\%$ depleted. The dust destruction process is not considered, as the charge of dust grains is too low ($q<5V$) to activate dust destruction ($q\approx100V$) in these AGN radiating regions.

In Figure~\ref{fig:4}, we present our dust-free AGN model and dusty AGN model built with four different AGN radiation fields on the standard optical diagnostic diagrams, compared with the optical data from SDSS DR16. Each model is presented by a grid consisting of constant metallicity and ionization parameter lines. 

As shown in Figure~\ref{fig:4}, both models provide very similar predictions at low metallicity ($12+\log\rm{(O/H)}<8.7$). However, at $12+\log\rm{(O/H)}>8.7$, the metallicity sensitive line ratios {[N~{\sc ii}]}/H$\alpha$, {[S~{\sc ii}]}/H$\alpha$, and {[O~{\sc i}]}/H$\alpha$ only vary a little in the dust-free models. In contrast, these line ratios continue to increase with metallicity in the dusty model. At $12+\log\rm{(O/H)}=9.0$, the predictions on {[N~{\sc ii}]}/H$\alpha$, {[S~{\sc ii}]}/H$\alpha$, and {[O~{\sc i}]}/H$\alpha$ from dust-free models are in general $0.3\,$dex lower than the the values from dusty models. At $12+\log\rm{(O/H)}=9.2$, the difference increases to $0.5\,$dex.

By comparing with the observed optical spectra of Seyfert galaxies, we find a significant fraction of Seyfert galaxies exceed the maximum {[N~{\sc ii}]}/H$\alpha$, {[S~{\sc ii}]}/H$\alpha$, and {[O~{\sc i}]}/H$\alpha$ values predicted by dust-free models, but still can be reasonably accounted for by the dusty models. For the Seyfert galaxies that have lower metallicity $12+\log\rm{(O/H)}<8.7$, both dust-free and dusty models give similar predictions for the model spectra. Given the fact that dust-free models fail to predict the observed {[N~{\sc ii}]}/H$\alpha$, {[S~{\sc ii}]}/H$\alpha$, and {[O~{\sc i}]}/H$\alpha$ ratios of Seyfert galaxies that have high metallicity $12+\log\rm{(O/H)}>8.7$, we conclude that dust depletion is likely to be a vital factor in ensuring reliable model at high metallicity regions and should be taken into account in the AGN model. Our finding agrees with \citet{dopita_are_2002} and \citet{groves_dusty_2004-1}.

\subsection{Abundance Set}\label{sec:4.2}

An abundance set provides information on the relative abundance for all primary nucleosynthetic elements (the lightest 30 elements) to the abundance of hydrogen. It usually consists of a standard abundance reference and abundance scaling relations characterizing how the relative abundance for the other 29 elements change with hydrogen when the total metallicity changes. The abundance set is a crucial input as it can directly affect the relative abundance of elements and thus change the emission line ratios predicted by AGN models. 

In the previous nebula photoionization models, a general solution for the abundances set is to adopt solar abundances as the standard abundance reference and linearly scale the abundance of all elements based on the total metallicity to solar metallicity ratio, except for helium and nitrogen. Although the sun is the star with the most accurate stellar atmosphere abundance measurements, solar abundances have their shortcomings: (i) there are elements that are not directly detected or only marginally detected in the solar atmosphere (F, Cl, Ne, Ar); (ii) He and Li are evolved in solar activity and thus their photosphere abundances do not necessarily reflect their protostar values; (iii) The abundance of oxygen, a crucial element for nebular physics, is difficult to measure in the solar spectrum, and the values provided by different measurements varied by a factor of $\sim2$ \citep{asplund_chemical_2009}.

\begin{figure*}[ht]
\epsscale{1.1}
\plotone{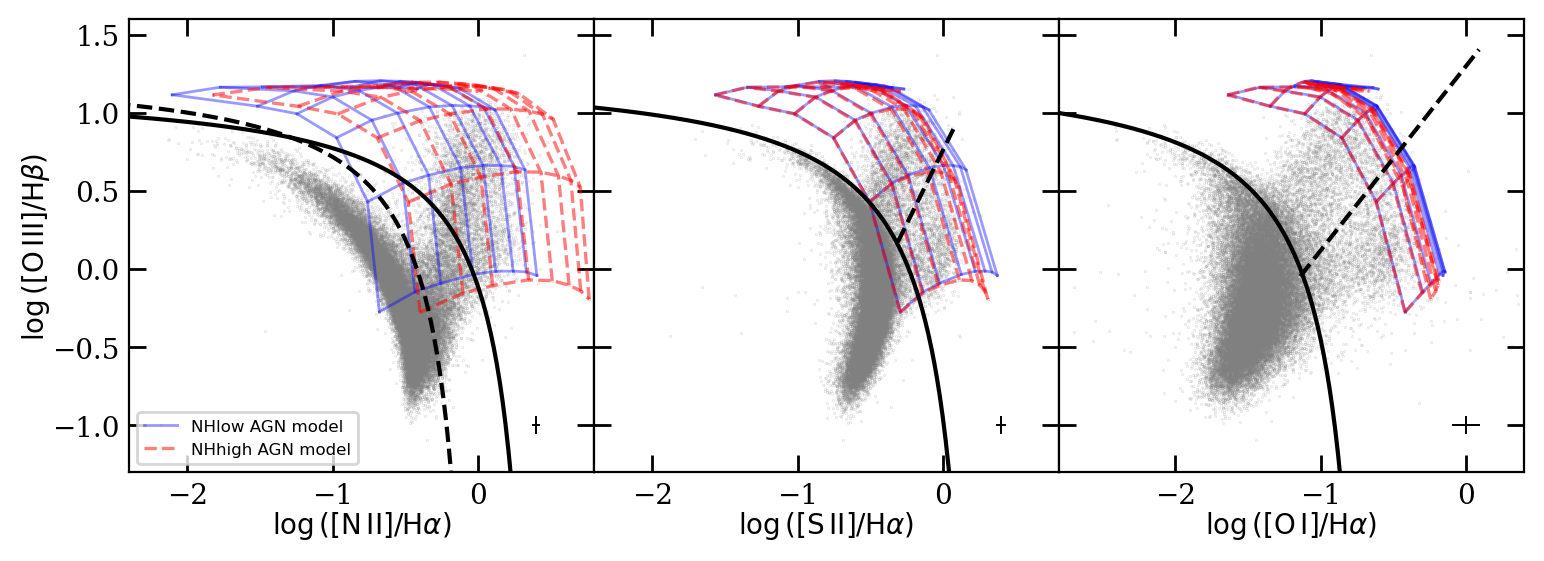}
\caption{Comparison of AGN models built with different N/O$-$O/H scaling relations (solid blue for `NHlow' and red dashed for `NHhigh') on the standard optical diagnostic diagrams. Each model is presented with constant metallicity lines (from left to right: $12+\log(\rm O/H)=8.28, 8.43, 8.54, 8.70, 8.80, 8.89, 8.96, 9.02$) and constant ionization parameter lines (from bottom to up: $\log(\rm U)= -3.8, -3.4, -3.0, -2.6, -2.2, -1.0$) at gas pressure $\log \rm(P/k)=7.4$. Both AGN models are dusty models with isobaric structure, non-uniform abundance sets, and $\log E_{\text{peak}}=-1.25$ OXAF radiation field calculated by MAPPINGS V.  Optical spectra for a large sample of galaxies from SDSS DR16 are presented with grey dots. The mean error bars are shown in the lower right corner. In all three panels, the black solid curves are the \citet{kewley_theoretical_2001} theoretical maximum starburst lines. In the left panel, the black dashed curve is the \citet{kauffmann_host_2003} empirical maximum starburst line. The black dashed lines in the right two panels are the Seyfert-LINER (low-ionization narrow emission-line regions) separation lines \citep{kewley_host_2006}.
\label{fig:5}}
\end{figure*}

Recently, \citet{2017MNRAS.466.4403N} provided a set of more realistic abundance scaling relations by studying the extensive stellar abundance data in the Milky Way. Instead of using solar abundance as a reference standard, they use the `Cosmic Abundance Standard' proposed by \citet{nieva_present-day_2012}, based on the average abundance over 29 local B stars in the Milky Way. In the `Cosmic Abundance Standard', \citet{nieva_present-day_2012} provides abundances for the eight most important nebular elements, He, C, N, O, Ne, Mg, Si, and Fe. For completeness of the standard abundance reference, \citet{2017MNRAS.466.4403N} extended the set to include the best estimation from meteoritic and the most recent solar abundances. Table 1 of \citet{2017MNRAS.466.4403N} shows the assembled standard abundance reference.

For abundance scaling relations, \citet{2017MNRAS.466.4403N} introduces a parametric enrichment factor $\zeta$ to characterize how atomic abundance scales with total abundance, which can be transferred into scales with a specific reference element such as oxygen or iron. With the extensive Milky Way stellar abundance data, they derived a set of non-uniform scaling relations for nebular modeling, which have been proved to significantly improve the behavior of HII models on the BPT diagram \citep{2017MNRAS.466.4403N,thomas_interrogating_2018}.

Despite the improvements in abundance scaling relations for most elements, disagreement on the N/O$-$O/H scaling relation remains among different modelers. The enrichment of nitrogen originates from two sources: a primary source that produces nitrogen through core-collapse supernovae in the native gas cloud \citep{van_zee_abundances_1998} and a secondary source from the delayed nucleosynthesis in hot-bottom burning and dredge-up in intermediate-mass stars \citep{renzini_advanced_1981}. While the primary source contributes to a constant ratio of nitrogen to oxygen (N/O) abundance with increasing oxygen abundance, the secondary nitrogen enrichment source starts to dominate from a particular oxygen abundance and leads to a rapidly growing N/O as oxygen abundance increases. 

Although consensus is reached on the two-phase nitrogen enrichment scenario, at what oxygen abundance the secondary source of nitrogen starts to dominate and how quickly it enriches the nitrogen abundance are still under discussion. There are two frequently used $\log$(N/O)$-\log$(O/H) relations in nebula modeling. We called them `NHlow' and `NHhigh' scaling relations based on their secondary nitrogen enrichment speed. The `NHlow' scaling relation is built with only stellar abundance data to avoid contamination from planetary nebulae that are hard to distinguish in low resolution data. The secondary component in the `NHlow' scaling relation ensures that the N/O ratio increases linearly with the oxygen abundance O/H. This results from nitrogen being a secondary element generated from the CNO cycle and its formation via the C and O already present in the star \citep{renzini_advanced_1981,van_zee_spectroscopy_1998}. By fitting only the stellar abundance data, a transition oxygen abundance between primary and secondary components at $12+\log(\text{O/H})\approx8.2$ is obtained \citep{groves_dusty_2004-1,gutkin_modelling_2016,2017MNRAS.466.4403N}. The `NHhigh' scaling relation is derived by adopting the best-fit $\log$(N/O)$-\log$(O/H) relation to nebula abundance data without assuming a fixed slope in advance, which results in a slope of 1.29 and a transition oxygen abundance at $12+\log(\text{O/H})\approx8.0$ \citep{dors_new_2017,perez-montero_bayesian-like_2019,carvalho_chemical_2020}. Both `NHlow' and `NHhigh' scaling relations are presented in the following equations, with equation (11) \citep{2017MNRAS.466.4403N} valid for $6.0\lesssim(12+\log(\rm O/H))\lesssim9.2$ and equation (12) \citep{carvalho_chemical_2020} valid for $8.0\lesssim(12+\log(\rm O/H))\lesssim9.0$.

\begin{small}
\begin{equation}\nonumber
\text{`NHlow'}:\log(\rm N/O)=\log(10^{-1.732}+10^{[\log(O/H)+2.19]})\eqno(11)
\end{equation}
\begin{equation}\nonumber
\text{`NHhigh'}:\log(\rm N/O)=1.29\times(12+\log(O/H))-11.84\eqno(12)
\end{equation}
\end{small}

Between the two N/O$-$O/H relations `NHlow' and `NHhigh', the difference in the N/O ratio at high metallicity end ($12+\log(\text{O/H})\approx9.0$) can be as large as $\sim0.4\,$dex, which will lead to a $\sim0.4\,$dex discrepancy in the predicted intensity of the nitrogen emission lines. This effect will pass on to line ratios that include nitrogen emission line, such as {[N~{\sc ii}]}$\,\lambda$6584/H$\alpha$ and {[N~{\sc ii}]}$\,\lambda$6584/{[O~{\sc ii}]}$\,\lambda$3727 which are frequently used as metallicity diagnostics. In Figure~\ref{fig:5}, we present two AGN models using all the same settings but the two different nitrogen scaling relations (`NHlow' and `NHhigh' respectively) on the standard optical diagnostic diagrams, compared with the optical spectra of Seyfert galaxies. As shown in Figure~\ref{fig:5}, the AGN model built with the `NHhigh' relation overall predicts higher {[N~{\sc ii}]}$\,\lambda$6584/H$\alpha$ than the AGN model built with `NHlow' relation. The discrepancy between the two models on {[N~{\sc ii}]}$\,\lambda$6584/H$\alpha$ is $\sim0.2\,$dex at $12+\log(\text{O/H})\approx8.0$, and increases to $\sim0.5$ dex when the metallicity reaches $12+\log(\text{O/H})\approx9.1$. As for {[S~{\sc ii}]}/H$\alpha$ and {[O~{\sc i}]}/H$\alpha$, the difference between two models are much smaller, at $<0.1\,$dex. 

Understanding the impact that different nitrogen abundance scaling relations can cause on nebula modeling is essential. Nevertheless, it is hard to decide on a uniform N/O$-$O/H relation for all AGN regions, as the N/O$-$O/H relation in the real environment can change for many reasons. For example, delayed nucleosynthesis in intermediate-mass stars in the star formation history of a galaxy can shift the transition oxygen abundance in N/O$-$O/H relation. The presence of Wolf-Rayet stars can also contribute to rapid nitrogen enrichment. Therefore, AGN regions with more Wolf-Rayet stars formed over their history can intrinsically have smaller transition oxygen abundances in their  N/O$-$O/H relations than other AGN regions.

Given the strong correlation between nitrogen emission line intensity and the secondary nucleosynthetic component of the N/O$-$O/H relation, as well as its dependence on the star formation history of the galaxy, we decide not to make an arbitrary assumption on which N/O$-$O/H relation is more suitable in AGN models. Instead, we adopt two versions of the AGN model that use `NHhigh' and `NHlow' N/O$-$O/H relations, respectively. As most of the stellar abundance data and nebula abundance lie between the `NHhigh' and `NHlow' relations, the predictions from the two versions of AGN models can place an upper limit and a lower limit for reference on the nitrogen emission line flux observed in AGNs. For readers who wish to use nitrogen emission lines to estimate gas metallicity based on theoretical strong line diagnostics, we caveat that the N/O$-$O/H relation in the model should be carefully determined in advance based on the class of objects to which the nitrogen-based diagnostics are applied. 

\subsection{Density Structure}

\begin{figure*}[hbt]
\epsscale{1.1}
\plotone{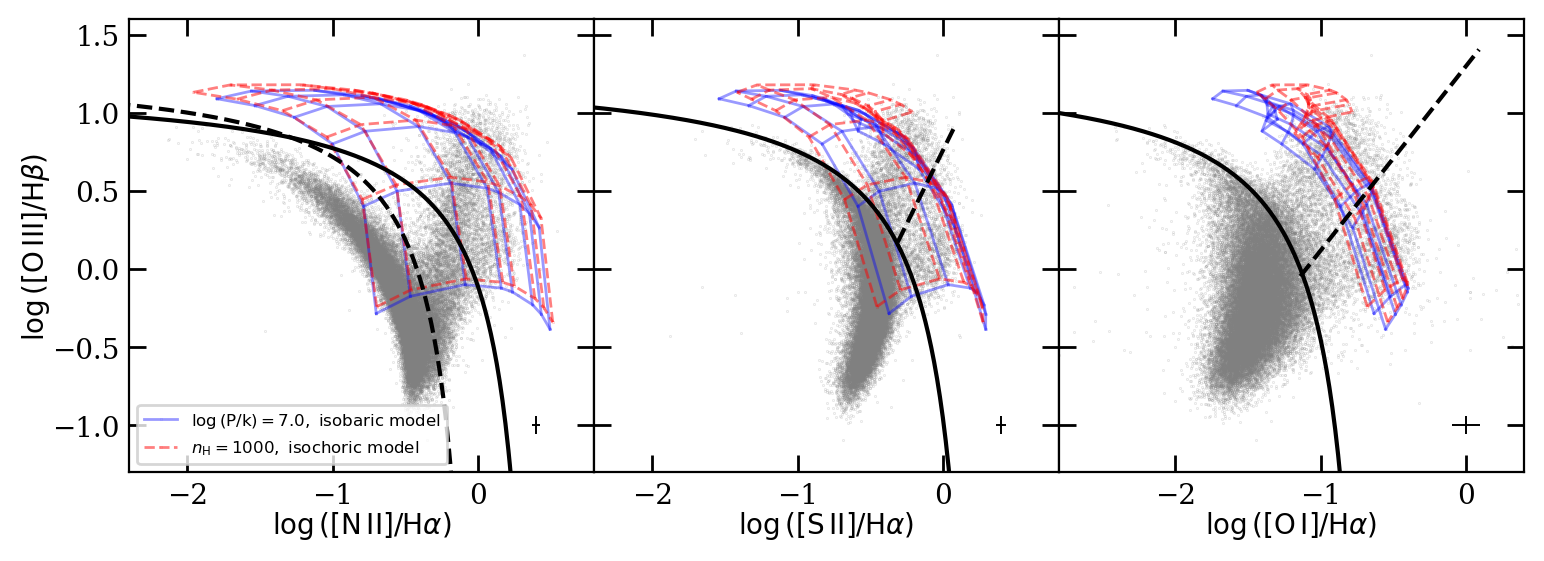}
\caption{Comparison of $\log E_{\text{peak}}=-1.5$ dusty AGN model using radiation pressure dominated isobaric structure and isochoric structure, respectively. Each model is presented with constant metallicity lines (from left to right: $12+\log(\rm O/H)=8.28, 8.43, 8.54, 8.70, 8.80, 8.89, 8.96, 9.02$) and constant ionization parameter lines (from bottom to up: $\log(\rm U)=-3.8, -3.4, -3.0, -2.6, -2.2, -1.8$) at gas pressure $\log \rm(P/k)=7.0$ (isobaric model, in solid blue) or at hydrogen density $n_{\rm H}=1000\rm cm^{-3}$ (isochoric model, in red dashed). `NHlow' nitrogen scaling relation is adopted in these models. Optical spectra for a large sample of galaxies from SDSS DR16 are presented with grey dots. The mean error bars are shown in the lower right corner.
\label{fig:6}}
\end{figure*}

A density structure must be chosen in nebula modeling to decide the model structure. One frequently used density structure is the isochoric model \citep[e.g.][]{feltre_nuclear_2016,carvalho_chemical_2020}, which assumes constant density in the photoionization region. However, constant-density photoionization models have difficulty reproducing the strong coronal lines observed in the NLRs of some Seyfert galaxies \citep[see][]{groves_dusty_2004}, which requires large ionization parameters and high radiation pressure environments \citep{binette_excitation_1996}. The assumption that NLRs have a constant density structure is also violated by observations. Using far-IR fine-structure lines for 170 local Seyfert galaxies, \citet{fernandez-ontiveros_far-infrared_2016} found evidence of the presence of a stratification of gas density in the emission regions of these galaxies.

The most successful solution to this problem by far is to adopt a radiative pressure-dominated isobaric model as the density structure \citep{dopita_are_2002}. The radiative pressure-dominated isobaric model can account for the presence of strong coronal lines and address the non-gravitational motions observed in the gas dynamics of the photoionized region within the NLR. In this model, the sum of radiative pressure and gas pressure remains constant within the photoionized region. It was suggested that using a radiative pressure-dominated isobaric structure in the photoionization model can provide more reasonable predictions than the standard isobaric and isochoric models for the emission-line spectra of the NLRs in Seyfert galaxies \citep{groves_dusty_2004,groves_dusty_2004-1}.

Figure~\ref{fig:6} presents the isochoric AGN model and the radiative pressure-dominated isobaric AGN model on the standard optical diagnostic diagrams to test their predictions with optical spectra of Seyfert galaxies. Both AGN models are dusty models with non-uniform abundance sets with `NHlow' scaling relation and a $\log E_{\text{peak}}=-1.5$ OXAF radiation field calculated by MAPPINGS V. With the assumption that the temperature in NLR is $\rm T\approx10^4\,K$, we set the gas pressure in the isobaric AGN model to be $\log \rm (P/k)=7.0$ and the hydrogen density in isochoric AGN model to be $n_{\rm H}=1000\,\rm cm^{-3}$ for a comparable test (given the definition of gas pressure $\rm P\sim n_{\rm H}kT$ in an ideal gas). 

As shown in Figure~\ref{fig:6}, the isochoric model and the radiative pressure-dominated isobaric model in general provide close predictions of [N~{\sc ii}]/H$\alpha$, [S~{\sc ii}]/H$\alpha$, [O~{\sc i}]/H$\alpha$, and [O~{\sc iii}]/H$\beta$, with discrepancy $<0.1$ dex. Exceptions are the theoretical ratios of [S~{\sc ii}]/H$\alpha$ and [O~{\sc i}]/H$\alpha$ when ionization parameter exceeds $\log(U)\approx-2.0$, where the isochoric AGN model predicts $\sim0.2-0.4$ dex higher ratios than radiative dominated isobaric AGN model. Here we conclude that both the isochoric model and radiative pressure-dominated isobaric model can provide reasonable and similar predictions on the standard optical diagnostic diagrams. The Comparison of `NHhigh' AGN models gives the same conclusion. Given that the isochoric model failed to explain the existence of strong coronal lines in NLRs, we adopt the radiative pressure-dominated isobaric model in the remainder of this paper.

\subsection{Gas Pressure}

\begin{figure*}[htb]
\epsscale{1.1}
\plotone{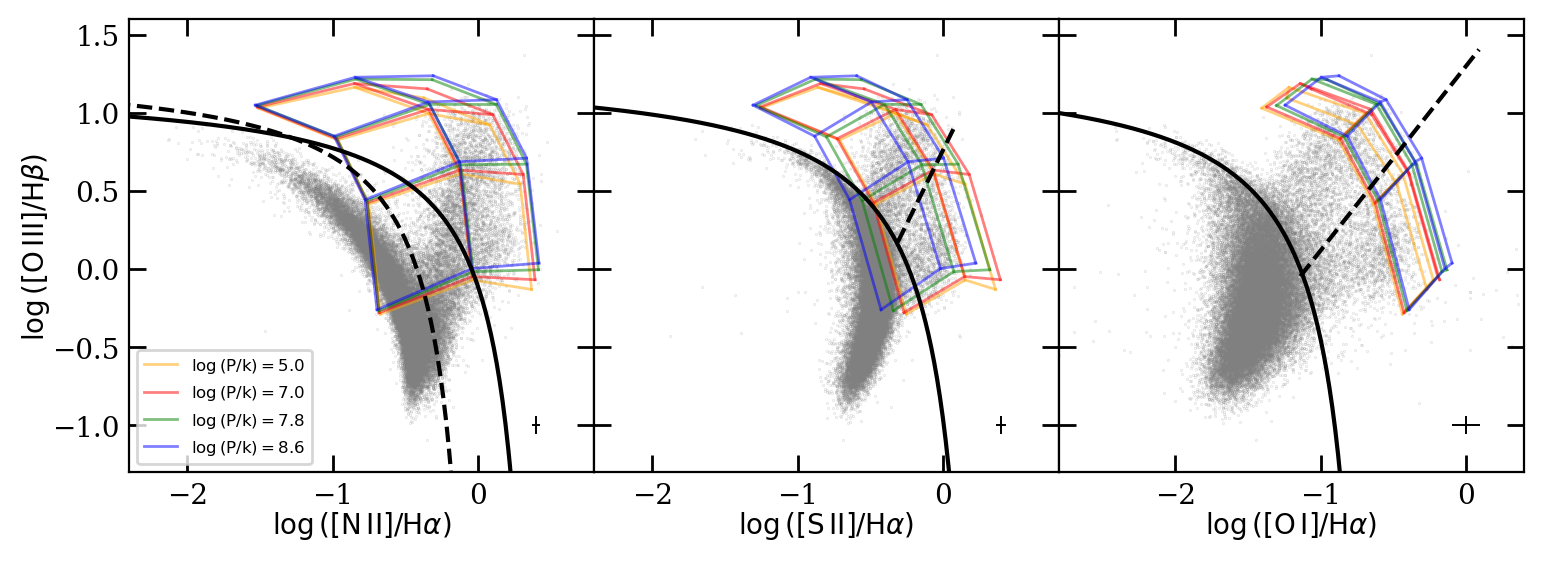}
\caption{Dusty $\log E_{\rm {peak}}=-1.25$ isobaric AGN models with different gas pressure $\log \rm(P/k)=$ 5.0, 7.0, 7.8, 8.6 shown in yellow, red, green, and blue grids on the standard optical diagnostic diagrams. Each model is presented with constant metallicity lines (from left to right: $12+\log(\rm O/H)=$8.28, 8.43, 8.70, 8.89, 9.02) and constant ionization parameter lines (from bottom to up: $\log(\rm U)=-3.8, -3.4, -3.0, -2.2$). `NHlow' nitrogen scaling relation is adopted in these models. Optical spectra for a large sample of galaxies from SDSS DR16 are presented with grey dots. The mean error bars are shown in the lower right corner.
\label{fig:7}}
\end{figure*}

The initial gas pressure is considered a preset parameter in the isobaric AGN model. Here we investigate how the value of initial gas pressure $\log (P_{\rm {gas}}/k)$ can affect the predictions on emission line ratios used in the standard optical diagnostic diagrams. As shown in Figure~\ref{fig:7}, at $\log (P_{\rm {gas}}/k)<7.0$, [N~{\sc ii}]/H$\alpha$, [S~{\sc ii}]/H$\alpha$, and [O~{\sc i}]/H$\alpha$ are independent of the initial gas pressure. At $\log (P_{\rm {gas}}/k)>7.0$, [S~{\sc ii}]/H$\alpha$ decrease $\sim 0.1$ dex when $\log (P_{\rm {gas}}/k)$ increase $\sim0.8$ dex, while [O~{\sc i}]/H$\alpha$ and [O~{\sc iii}]/H$\beta$ increase at a similar rate. 

The emission line ratios used in the standard optical diagnostic diagrams have a much stronger dependence on metallicity and ionization parameters than the initial gas pressure, which is an essential property of metallicity and ionization parameter diagnostics. Nevertheless, the effect from gas pressure should be considered at $\log (P_{\rm {gas}}/k)>7.0$, as a $\sim 0.8$ dex change in $\log (P_{\rm {gas}}/k)$ can lead to $\sim 0.1$ dex or even more variation in the predicted emission line ratios (see Zhu et al. (2023b, in prep) for more discussion). 

\section{Z-Q DIAGNOSTIC DIAGRAM}\label{sec:dia_diag}

Once the dust properties, element abundance, and density structure are determined in photoionization models, the emission-line spectra in the NLR of AGNs are directly related to the metallicity, ionization parameter, and gas pressure in the photoionized region as well as the shape of the ionizing radiation field. For star-forming regions, it was found that some emission line ratios present exclusive dependence on the gas metallicity ([N~{\sc ii}]/[O~{\sc ii}]), ionization parameter ([O~{\sc iii}]/[O~{\sc ii}]) and gas pressure ([S~{\sc ii}]/[S~{\sc ii}]), making them good diagnostics for observations interpretations (see \citet{kewley_understanding_2019} for a review). For AGN regions, it is also suggested that [N~{\sc ii}]/H$\alpha$ \citep{carvalho_chemical_2020} and [N~{\sc ii}]/[O~{\sc ii}] \citep{castro_new_2017} present strong dependence on gas metallicity. The [O~{\sc iii}]/[O~{\sc ii}] ratio is also proposed as an ionization indicator by \citet{groves_dusty_2004-1}. 

We construct AGN diagnostic diagrams with metallicity-sensitive (Z) line ratios and ionization-sensitive (Q) line ratios as the x and y axes, respectively.

This section examines our selected AGN model on the Z-Q diagnostic diagrams. We also extend our examination to infrared and UV wavelengths. Here we only present diagnostic diagrams that have observational data to compare against, to provide a test of the models. By comparing with the observations of Seyfert galaxies in Section~\ref{sec:sample}, we test the predictions of our AGN model across a broad range of wavelengths. HII models and observations of star-forming galaxies are also presented (when available) for searching potential AGN-HII separation diagrams.

\subsection{Optical Diagnostic Diagrams}\label{sec:5.1}

\begin{figure*}[htb]
\epsscale{0.7}
\plotone{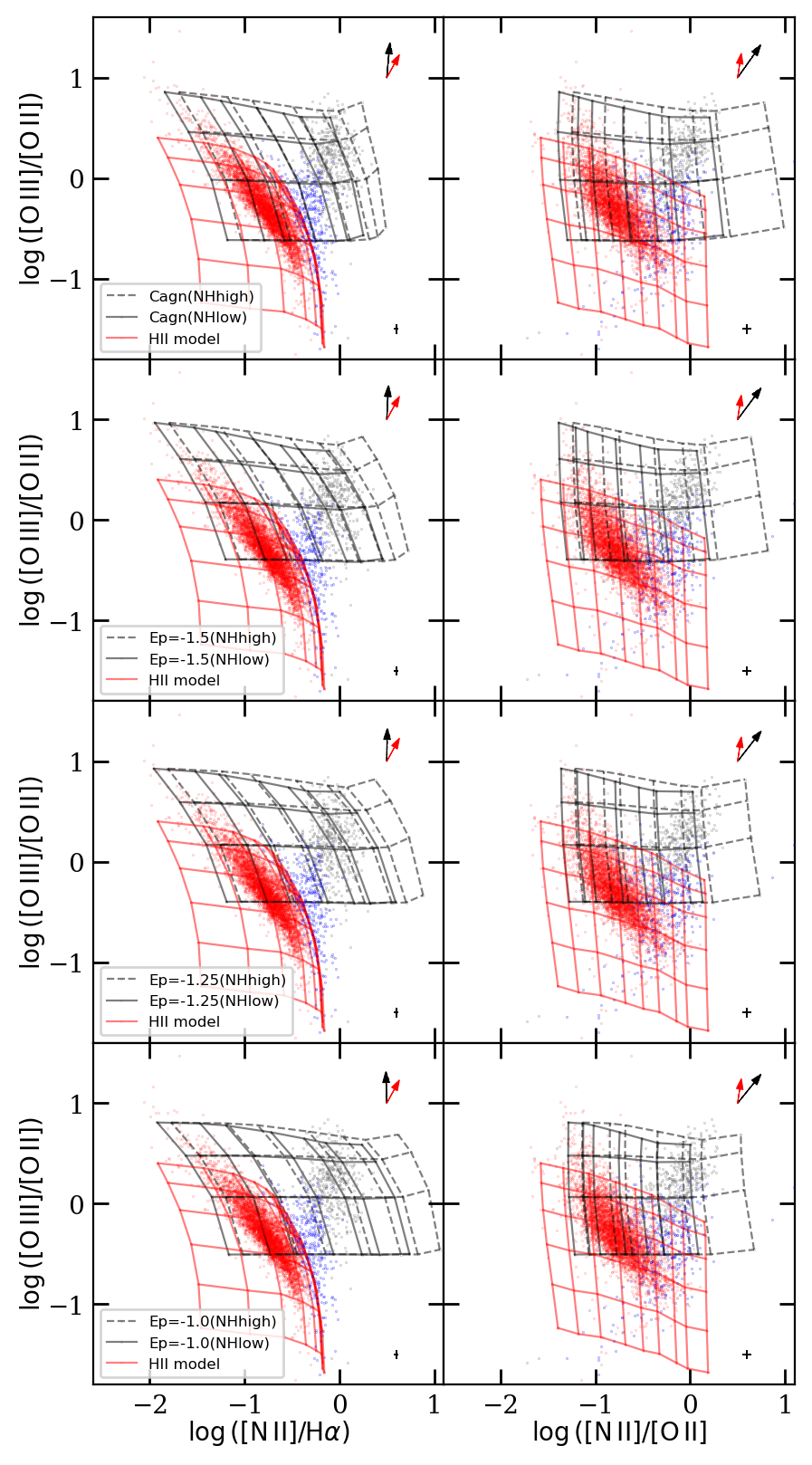}
\caption{Optical diagnostic diagrams for AGN regions and HII regions. Dusty, isobaric AGN models with different N/O$-$O/H scaling relations (`NHlow' in solid lines and `NHhigh' in dash lines), different radiation fields (from top to bottom panels: `Cagn', $\log E_{\text{peak}}= -1.5, -1.25, -1.0$) are presented in black grids that consist of constant metallicity lines (from left to right: $12+\log(\rm O/H)=8.00, 8.28, 8.43, 8.70, 8.89, 9.02, 9.26$) and constant ionization parameter lines (from bottom to top: $\log(\rm U)=-3.4, -3.0, -2.6, -2.2$) at gas pressure $\log \rm(P/k)=7.4$. The black arrow in the upper right corner represents the average effect of gas pressure when it changes from $\log \rm(P/k)=7.4$ to $\log \rm(P/k)=8.2$. HII model is shown with a red grid that consists of constant metallicity lines (from left to right: $12+\log(\rm O/H)=7.75, 8.15, 8.43, 8.63,  8.76, 8.85, 8.94, 9.00, 9.10$) and constant ionization parameter lines (from bottom to top: $\log(\rm U)=-3.75, -3.5, -3.25, -3.0, -2.75, -2.5$) at gas pressure $\log \rm(P/k)=5.4$. The red arrow in the upper right corner represents the average effect of gas pressure when it changes from $\log \rm(P/k)=5.4$ to $\log \rm(P/k)=7.4$. Optical spectra for galaxies with S/N$\geq$5 from SDSS DR16 are presented with colors reflecting their classifications: red dots stand for HII galaxies, black dots stand for Seyfert galaxies, and blue dots stand for composite galaxies. The mean error bars for all galaxies are shown in the lower right corner.\label{fig:8}}
\end{figure*}

\begin{figure*}[htb]
\epsscale{0.7}
\plotone{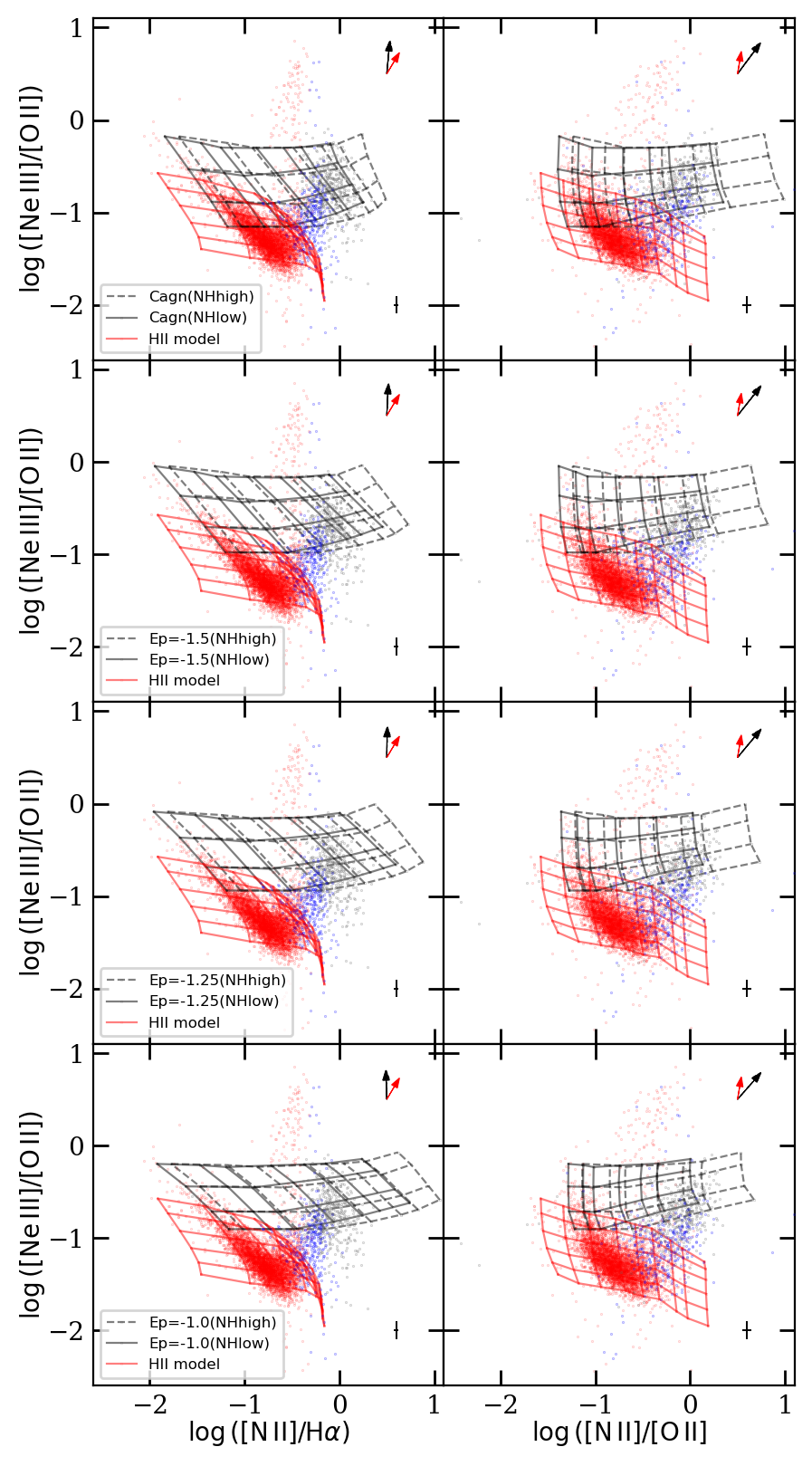}
\caption{Same as Figure~\ref{fig:8} but adopting {[Ne~{\sc iii}]}/{[O~{\sc ii}]} as the ionization parameter diagnostic. \label{fig:11}}
\end{figure*}

We present a set of diagnostic diagrams in Figure~\ref{fig:8} and Figure~\ref{fig:11} to examine our AGN models and their potential to determine the metallicity and ionization properties of AGN. We present AGN models built with the four different radiation fields introduced in Section~\ref{sec:3.1}, together with the HII model, to show the difference between these two mechanisms. We also consider the effect of gas pressure by adding an arrow on the diagram to present the impact on both line ratios introduced by the change in gas pressure. As discussed in Section~\ref{sec:4.2}, AGN models built with `NHlow' and `NHhigh' N/O$-$O/H scaling relations are presented in each diagram. Optical spectra for galaxies in SDSS DR16 are presented for model testing, applying a higher S/N ratio (S/N$\geq$5) to reduce the scatter in the {[Ne~{\sc iii}]}/{[O~{\sc ii}]} ratio.

As shown in Figure~\ref{fig:8}, the AGN models generally accommodate on the optical Seyfert spectra for the [NII]/Ha and [NII]/[OII] ratios. By adjusting gas pressure, the AGN models can predict those data lying outside the upper right corner of our model grids. We also notice some Seyfert observations that can only be predicted by the `NHhigh' AGN model. These could be the Seyferts that intrinsically have higher nitrogen-to-oxygen abundance. This hypothesis can be tested with measurements of their star formation history.

Figure~\ref{fig:8} and Figure~\ref{fig:11} reveal that the change in radiation field has a more significant impact on metallicity-sensitive line ratios ($\sim0.5$ dex at the high metallicity end) than ionization parameter-sensitivity line ratios ($\sim$0.1 dex for [OIII]/[OII], $\sim0.3$ dex for [NeIII]/[OII] at low ionization parameter end). The effect of gas pressure changing from $\log \rm(P/k)=7.0$ to $\log \rm(P/k)=8.2$ is $\lesssim0.3$ dex. Last but not least, the effect of the N/O abundance ratio can be as large as $\sim0.6$ dex on {[N~{\sc ii}]}/{[O~{\sc ii}]}. This effect becomes smaller for {[N~{\sc ii}]}/H$\alpha$ ($\sim0.3$ dex) and can be neglected for other line ratios that do not include nitrogen lines. These effects highlight the importance of parameter degeneracy. Parameter degeneracy should be carefully reviewed when applying metallicity and ionization parameter diagnostics for AGN regions (see Paper II for further discussion).

\subsection{Ultraviolet Diagnostic Diagrams}

\begin{figure*}[htb] 
\epsscale{0.9}
\plotone{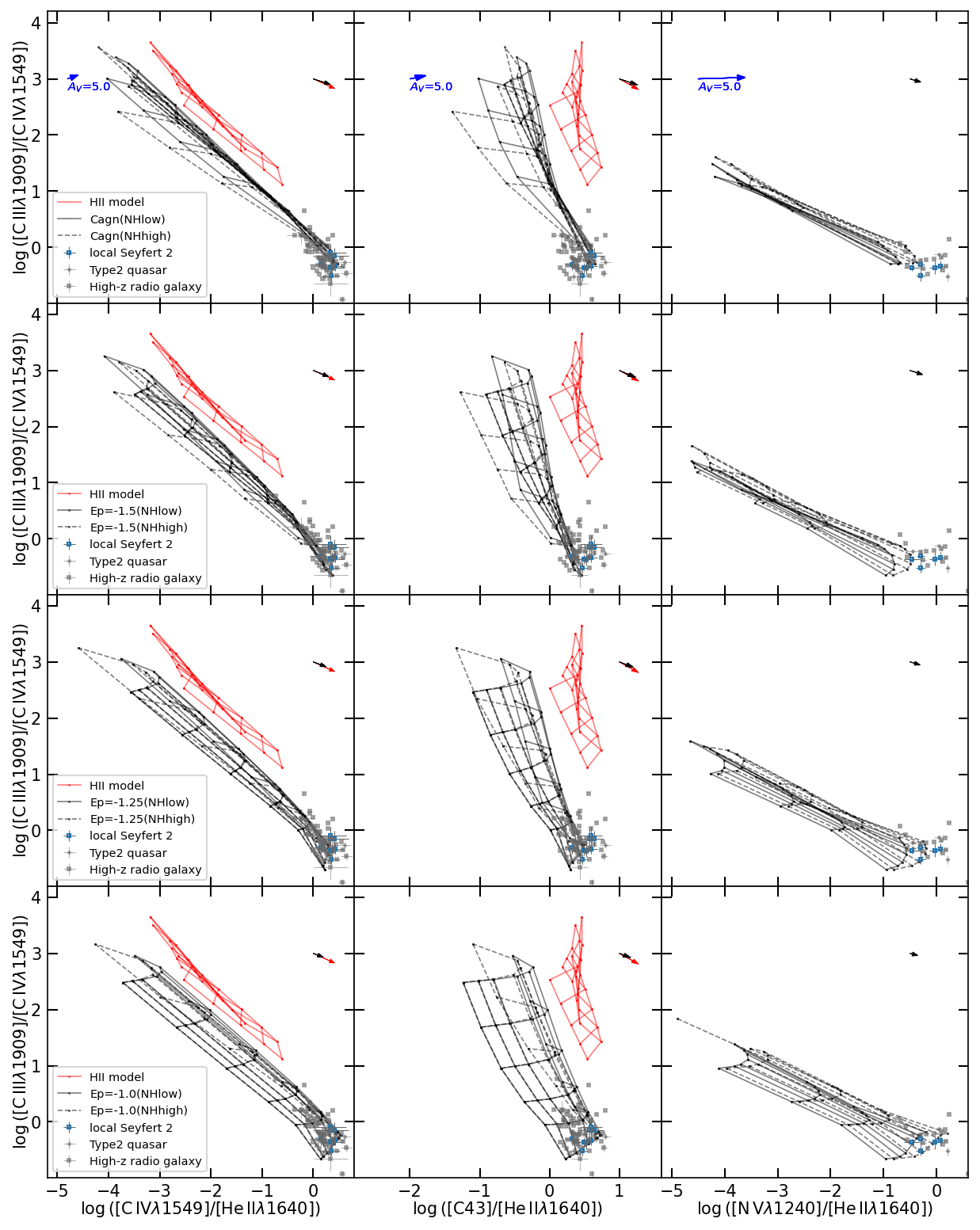}
\caption{UV diagnostic diagrams for AGN regions and HII regions. Dusty, isobaric AGN models with different N/O$-$O/H scaling relations (`NHlow' in solid lines and `NHhigh' in dash lines), different radiation fields (from top to bottom panels: `Cagn', $\log E_{\text{peak}}= -1.5, -1.25, -1.0$) are presented in black grids that consist of constant metallicity lines (from left to right: $12+\log(\rm O/H)=8.00, 8.28, 8.43, 8.70, 8.89, 9.02, 9.26$) and constant ionization parameter lines (from top to bottom: $\log(\rm U)=-3.8, -3.4, -3.0, -2.6, -2.2, -1.0$) at gas pressure $\log \rm(P/k)=7.4$. The black arrow in the upper right corner represents the average effect of gas pressure when it changes from $\log \rm(P/k)=7.4$ to $\log \rm(P/k)=8.2$. HII model is shown with a red grid that consists of constant metallicity lines (from left to right: $12+\log(\rm O/H)=7.75, 8.15, 8.43, 8.63, 8.76, 8.85, 8.94, 9.00, 9.10$) and constant ionization parameter lines (from top to bottom: $\log(\rm U)=-3.5, -3.25, -3.0, -2.75, -2.5$) at gas pressure $\log \rm(P/k)=5.4$. The red arrow in the upper right corner represents the average effect of gas pressure when it changes from $\log \rm(P/k)=5.4$ to $\log \rm(P/k)=7.4$. UV spectra for local Seyfert 2 galaxies, Type2 quasars, and high$-z$ radio galaxies are collected from \citet{dors_semi-empirical_2019} and presented with hollow blue squares, solid grey dots, and solid grey squares. The blue arrows at the upper-left corner in the top three diagrams indicate the $A_V=5.0\,$mag dust extinction correction for the UV observations.
\label{fig:9}}
\end{figure*}

Observations at the ultraviolet (UV) wavelengths are becoming more observable in the distant universe, especially with JWST, as a result of rest-frame UV emission lines being shifted to the infrared. Because shocks generally present stronger low excitation UV forbidden lines than HII regions. AGN UV diagnostic diagrams are potential tools for excitation source separation between star-forming regions, AGN regions, and shocks \citep{groves_dusty_2004-1,sutherland_effects_2017,dopita_spectral_1995}.

At UV wavelengths, three frequently used strong line diagnostics are {[C~{\sc iv}]}$\,\lambda\,$1549/{[C~{\sc iii}]}$\,\lambda\,$1909  (an ionization parameter diagnostic), as well as {[C~{\sc iv}]}$\,\lambda\,$1549/{[He~{\sc ii}]}$\,\lambda\,$1640 and {[C~{\sc iii}]}$\,\lambda\,$1909/{[He~{\sc ii}]}$\,\lambda\,$1640 (metallicity diagnostics). These three line ratios were first explored by \citet{villar-martin_ionization_1997} in the case of high$-z$ radio galaxies. Here we adopt the {[C~{\sc iv}]}/{[C~{\sc iii}]}$-${[C~{\sc iv}]}/{[He~{\sc ii}]} diagram, replace the {[C~{\sc iii}]}/{[He~{\sc ii}]} line ratio with C43/{[He~{\sc ii}]}\citep[suggested by][]{dors_semi-empirical_2019} (C43 is the sum of {[C~{\sc iii}]}$\,\lambda\,$1909 and {[C~{\sc iv}]}$\,\lambda\,$1549), and add another metallicity sensitive line ratio {[N~{\sc v}]}$\,\lambda\,$1240/{[He~{\sc ii}]}$\,\lambda\,$1640  \citep[suggested by][]{nagao_gas_2006} to build our UV diagnostic diagram, as presented in Figure~\ref{fig:9}. We explore AGN models with different radiation fields, the effect of gas pressure, and the impact of different N/O$-$O/H scaling relations. The HII model is also presented when applicable. For comparison with observations, we also include the UV observations introduced in Section~\ref{sec:2.2}.

In general, our AGN models can predict the observations of Seyfert galaxies both nearby and in the high$-z$ universe. It is striking that almost all Seyfert data lie at the lower right corner of AGN model grids, which suggests that these data, on average, have much higher ionization parameters ($\log(U)\approx -1.0$) than those observed in SDSS optical data ($\log(U)\lesssim -2.0$). This phenomenon is consistent with the findings of the previous studies \citet{nagao_gas_2006} and \citet{dors_semi-empirical_2019}. Among the 9 local Seyfert galaxies in our UV sample, 7 have IR observations in \citet{fernandez-ontiveros_far-infrared_2016} and 5 have optical observations in the SDSS DR14 or the Siding Spring Southern Seyfert Spectroscopic Snapshot Survey (S7) database \citep{dopita_probing_2015-1,thomas_probing_2017}. For these galaxies, observations at infrared and optical wavelengths are consistent with UV observations, suggesting a high ionization parameter ($\log(U)\approx -1.0$) by comparing with the IR and optical diagnostics we developed based on our AGN model (for diagnostics, see Zhu et al. 2023b, in prep). We also find that all 7 galaxies that have IR observations are classified as `S1h' in \citet{fernandez-ontiveros_far-infrared_2016} instead of `S2', due to the presence of broad line components in their IR spectra. This suggests that the higher ionization parameters of these Seyfert galaxies could be a result of being contaminated by the emissions from broad-line regions. Further information is needed to determine what causes this distinction between optical and UV observations. Specifically, galaxies with both optical and UV data covering a broad wavelength range are needed to test the reliability of these new UV diagnostics.

Figure~\ref{fig:9} shows that the AGN model built with the hardest radiation field ($\log E_{\text{peak}}=$-1.0) performs better in predicting most of the observations than other $E_{\text{peak}}$ values. This suggests that, if not all, most of the observed AGNs have relatively hard radiation fields. We also note that the effect of gas pressure on these UV line ratios is less ($\lesssim 0.2$ dex) than in the optical band. The impact of different N/O$-$O/H scaling relations is mostly on the {[N~{\sc v}]}/{[He~{\sc ii}]} ratio, where a significant fraction of the observed Seyferts can only be predicted by the `NHhigh' AGN model.

Together with the HII model, we find that both {[C~{\sc iv}]}/{[C~{\sc iii}]}$-${[C~{\sc iv}]}/{[He~{\sc ii}]} diagram and C43/{[He~{\sc ii}]}$-${[C~{\sc iv}]}/{[He~{\sc ii}]} diagram have well separated HII and AGN models. Optical and UV spectroscopic observations of AGN, and UV observations for HII regions are needed to test the ability of these diagrams to serve as excitation diagnostic tools at UV wavelengths.

\subsection{Infrared Diagnostic Diagrams}
	\label{sec:ob_comp}

\begin{figure*}[htb]
\epsscale{0.9}
\plotone{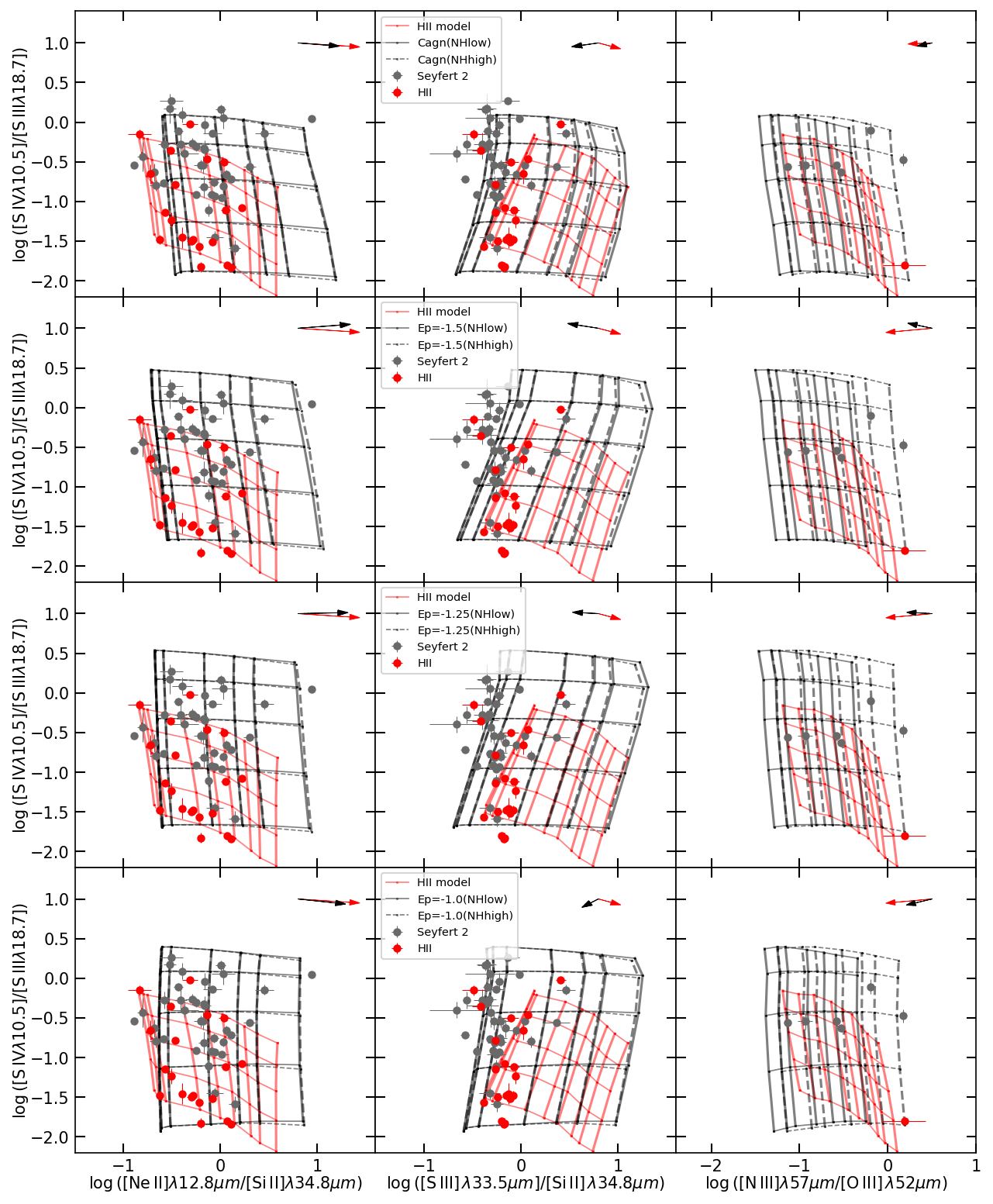}
\caption{Infrared diagnostic diagrams for AGN regions and HII regions. Dusty, isobaric AGN models with different N/O$-$O/H scaling relations (`NHlow' in solid lines and `NHhigh' in dash lines), different radiation fields (from top to bottom panels: `Cagn', $\log E_{\text{peak}}= -1.5, -1.25, -1.0$) are presented in black grids that consist of constant metallicity lines (from left to right: $12+\log(\rm O/H)=8.00, 8.28, 8.43, 8.70, 8.89, 9.02, 9.26$) and constant ionization parameter lines (from bottom to top: $\log(\rm U)=-3.8, -3.4, -3.0, -2.6, -2.2$) at gas pressure $\log \rm(P/k)=7.4$. The black arrow in the upper right corner represents the average effect of gas pressure when it changes from $\log \rm(P/k)=7.4$ to $\log \rm(P/k)=8.2$. HII model is shown with a red grid that consists of constant metallicity lines (from left to right: $12+\log(\rm O/H)=7.75, 8.15, 8.43, 8.63, 8.76, 8.85, 8.94, 9.00, 9.10$) and constant ionization parameter lines (from bottom to top: $\log(\rm U)=-3.5, -3.25, -3.0, -2.75, -2.5$) at gas pressure $\log \rm(P/k)=5.4$. The red arrow in the upper right corner represents the average effect of gas pressure when it changes from $\log \rm(P/k)=5.4$ to $\log \rm(P/k)=7.4$. Infrared spectra for Seyfert 2 galaxies and HII regions are collected from \citet{fernandez-ontiveros_far-infrared_2016} and presented with grey circles and red circles for comparison. \label{fig:10}}
\end{figure*}

To obtain reasonable predictions at infrared wavelengths, dust should be carefully treated and accounted for in AGN models, which a dust-free model can never do. After a broad investigation of the line ratios in the infrared, we select {[S~{\sc iv}]}$\,\lambda\,$10.5$\rm \mu m$/{[S~{\sc iii}]}$\,\lambda\,$18.7$\rm \mu m$ as a potential ionization parameter sensitive line ratio, together with three metallicity sensitive line ratios {[Ne~{\sc ii}]}$\,\lambda\,$12.8$\rm \mu m$/{[Si~{\sc ii}]}$\,\lambda\,$34.8$\rm \mu m$, {[S~{\sc iii}]}$\,\lambda\,$33.5$\rm \mu m$/{[Si~{\sc ii}]}$\,\lambda\,$34.8$\rm \mu m$, and {[N~{\sc iii}]}$\,\lambda\,$57$\rm \mu m$
/{[O~{\sc iii}]}$\,\lambda\,$52$\rm \mu m$ to build our infrared diagnostic diagrams, as presented in Figure~\ref{fig:10}. Similar to the study in optical and UV bands, we explore AGN models with different radiation fields, the effect of gas pressure, and the impact of different N/O$-$O/H scaling relations. The HII-region models are also presented. For comparison, we collect the latest infrared observations via Spitzer for Seyfert 2 galaxies and HII regions (See the introduction in Section~\ref{sec:2.3}).
 
As shown in Figure~\ref{fig:10}, we find that AGN models built with OXAF radiation fields ($\log E_{\text{peak}}=-1.5, -1.25, -1.0$) can all predict most of the Seyfert 2 observations on the {[S~{\sc iv}]}/{[S~{\sc iii}]} and {[Ne~{\sc ii}]}/{[Si~{\sc ii}]} diagram. On the {[S~{\sc iv}]}/{[S~{\sc iii}]} and {[S~{\sc iii}]}/{[Si~{\sc ii}]} diagram, observations of Seyfert 2 galaxies are best fitted by AGN models with the hardest radiation field ($\log E_{\text{peak}}=-1.0$). All four AGN models can reasonably predict Seyfert 2 galaxies on the {[S~{\sc iv}]}/{[S~{\sc iii}]} and {[N~{\sc iii}]}/{[O~{\sc iii}]} diagram. But the lack of observations on the {[N~{\sc iii}]}/{[O~{\sc iii}]} ratios prevents us from using this diagram for model examination. Overall, we conclude that the AGN model with $\log E_{\text{peak}}=-1.0$ is the best-fit model according to the infrared diagnostic diagrams. The AGN model with the CLOUDY default radiation field systematically predicts a lower {[S~{\sc iv}]}/{[S~{\sc iii}]} ratio compared to both observations and other AGN models.

At infrared wavelengths, a harder AGN radiation field does not always result in higher emission line ratios, such as {[S~{\sc iv}]}/{[S~{\sc iii}]} and {[S~{\sc iii}]}/{[Si~{\sc ii}]} in the diagrams. Despite the non-monotone dependence, the change in these line ratios introduced by different AGN radiation fields is at most $\lesssim0.5$ dex. The other two line ratios are almost independent of the hardness of the AGN radiation field. The effect of gas pressure is of similar importance, which can change all metallicity-sensitive line ratios up to $\sim0.3$ dex. The impact from different N/O$-$O/H scaling relations is only present in the nitrogen-based line ratio.

Together with the HII model, we find that both AGN and HII models match relatively well with the observed data. However, because of the considerable overlap between the two models on these diagrams, it is hard to distinguish HII regions and AGN regions simply from these infrared diagnostic diagrams. Further exploration of theoretical models and multi-wavelength data covering a broader wavelength range sets are required to identify a well-performing HII$-$AGN separation diagram in the infrared. 

\section{Discussion}\label{sec:disc}

With the aim of obtaining a reliable AGN model for emission line studies, we have explored a wide range of factors that can affect the predictions of AGN models, including AGN radiation fields, photoionization code, the presence of dust, element abundance, model density profile, as well as gas pressure, metallicity, and ionization parameter. By comparing the model predictions with observations at UV, optical, and IR wavelengths, we show that the dusty, radiation pressure-dominated isobaric AGN model provides the best prediction of observed Seyfert galaxies, with some scatter caused by the non-uniform hardness of the AGN radiation field and different nitrogen to oxygen abundance ratios. 

We use our best-fit AGN models with HII-region models to explore the model behavior on UV, optical, and infrared diagnostic diagrams where we have observational constraints from data in the literature. We find that the majority of AGN and HII models match remarkably well with observations in optical, UV, and IR wavelengths. We find two potential HII$-$AGN separation diagrams in the UV band, the {[C~{\sc iv}]}/{[C~{\sc iii}]}$-${[C~{\sc iv}]}/{[He~{\sc ii}]} and C43/{[He~{\sc ii}]}$-${[C~{\sc iv}]}/{[He~{\sc ii}]} diagrams, which require further testing with future data.

AGN modeling is essential for theoretical metallicity diagnostics for AGN regions. Studies on AGN diagnostics are relatively lacking compared to the rich investigations on strong line diagnostics for HII regions \citep[for published AGN metallicity diagnostics, see][]{storchi-bergmann_chemical_1998,nagao_gas_2006,feltre_nuclear_2016,castro_new_2017,carvalho_chemical_2020}. The main difficulty for an extensive diagnostics search is a consistent AGN model that can provide reasonable predictions across a broad range of wavelengths. In a future paper, we will investigate the use of our AGN models to determine the metallicity in AGN galaxies.

The AGN model also enables the study of mixing mechanisms in composite galaxies. While researchers usually classify galaxies into star-forming galaxies, Seyferts, and LINERs based on the standard optical diagnostic diagrams, some galaxies contain more than one excitation mechanism. For example, NGC\,253 and NGC\,1365 are found to present both star formation and AGN activities by \citet{sharp_three-dimensional_2010}. In 2014, \citet{davies_starburstagn_2014} confirmed NGC\,7130 as a starburst-AGN mixed galaxy and estimated that AGN activity contributes $\sim35\%$ of the H$\alpha$ luminosity in this galaxy. Galaxies with mixing excitation sources are also identified in \citet{rich_galaxy-wide_2011,rich_composite_2014,dagostino_comparison_2019,cairos_warm_2022}.
 
For composite galaxies, instead of arbitrarily assigning them as star formation dominated galaxies or AGN dominated galaxies or others, it is more valuable to estimate the emission line contributions from star-forming excitation, AGN excitation, and other excitation mechanisms. For example, the H$\alpha$ emission line is frequently used in determining the star formation rate (SFR) of a galaxy. However, as we can see from our AGN model, AGN excitation can also contribute to H$\alpha$ emission, which can significantly affect the SFR measurement in a composite galaxy (up to $40\%$ in NGC~7130 and NGC~2410; \citet{davies_starburstagn_2014,davies_starburstagn_2014-1}) even if it is dominated by star formation. More recently, \citet{cairos_warm_2022} used standard optical diagnostic diagrams and found $\sim37\%$ H$\alpha$ luminosity in the blue compact galaxy Haro 14 is contributed by mechanisms other than OB-star photoionization, which will overestimate the total star formation rate by a factor of $\sim1.6$. This mixing effect will occur whenever emission lines are used to interpret the properties of composite galaxies. 

The most popular methods used to calculate AGN and star-forming contributions are identifying an AGN-star formation mixing sequence on the BPT diagram. With the pure AGN basis point and pure star formation basis point empirically selected on the BPT diagram, a mixing sequence can then be built to interpret the fractions of different mechanisms for composite galaxies based on their location on the mixing sequence \citep[e.g.][]{wild_timing_2010,davies_dissecting_2016,husemann_close_2019,molina_enhanced_2023}. However, as pointed out by \citet{davies_dissecting_2016}, for galaxies that only have mixed spectra and without `pure' spectra, the identification of mixing sequence is arbitrary and the measurement of excitation fractions could have a systematic offset and a larger scatter.

Fortunately, a reliable AGN model can help to solve this problem. Using the AGN model together with the HII model, a theoretical mixing sequence can be produced in the standard optical diagnostic diagrams or other diagnostic diagrams \citep[for example, see][]{dagostino_starburstagn_2018}. With the pure basis points determined by theoretical models, the theoretical mixing sequence can provide more consistent predictions for a large sample of galaxies, whether or not they contain `pure' spectra. 

Furthermore, with theoretical AGN and HII models, we will be able to build mixing sequences in the UV and infrared diagnostic diagrams, enabling mixing excitations separations for observations at these wavelengths, which will significantly benefit high-$z$ studies in the JWST era. To test the reliability of mixing sequences at UV and IR wavelength, long-wavelength coverage multi-wavelength spectroscopy for the same HII regions and AGN is needed.

\section{Acknowledgement}

Parts of this research were conducted by the Australian Research Council Centre of Excellence for All Sky Astrophysics in 3 Dimensions (ASTRO 3D), through project number CE170100013. L.J.K. gratefully acknowledges the support of an ARC Laureate Fellowship (FL150100113).

Funding for the Sloan Digital Sky Survey V has been provided by the Alfred P. Sloan Foundation, the Heising-Simons Foundation, the National Science Foundation, and the Participating Institutions. SDSS acknowledges support and resources from the Center for High-Performance Computing at the University of Utah. The SDSS website is www.sdss5.org.

SDSS is managed by the Astrophysical Research Consortium for the Participating Institutions of the SDSS Collaboration, including the Carnegie Institution for Science, Chilean National Time Allocation Committee (CNTAC) ratified researchers, the Gotham Participation Group, Harvard University, The Johns Hopkins University, L’Ecole polytechnique fédérale de Lausanne (EPFL), Leibniz-Institut für Astrophysik Potsdam (AIP), Max-Planck-Institut für Astronomie (MPIA Heidelberg), Max-Planck-Institut für Extraterrestrische Physik (MPE), Nanjing University, National Astronomical Observatories of China (NAOC), New Mexico State University, The Ohio State University, Pennsylvania State University, Smithsonian Astrophysical Observatory, Space Telescope Science Institute (STScI), the Stellar Astrophysics Participation Group, Universidad Nacional Autónoma de México, University of Arizona, University of Colorado Boulder, University of Illinois at Urbana-Champaign, University of Toronto, University of Utah, University of Virginia, and Yale University.

\bibliography{PaperI.bib}{}
\bibliographystyle{aasjournal}

\end{document}